\documentclass[journal]{IEEEtran}
\usepackage{graphicx}
\usepackage[table,xcdraw]{xcolor}
\usepackage{amsmath}
\usepackage{amssymb}
\usepackage{booktabs}
\usepackage{multirow}
\usepackage{color}
\usepackage{adjustbox}
\usepackage{xcolor}
\usepackage{cite}
\usepackage{float}
\usepackage{algorithm}
\usepackage[vlined, ruled, algo2e]{algorithm2e}
\usepackage{csquotes}

\usepackage[inline]{enumitem}
\usepackage{hyperref}
\usepackage{varwidth}

\ifodd 1
\newcommand{\revtwo}[1]{{\color{black} #1}} 

\else

\fi

\begin{document}

\title{A Survey of Data Fusion in Smart City Applications}

\author{{Billy Pik Lik Lau, Sumudu Hasala Marakkalage, Yuren Zhou, Naveed Ul Hassan, Chau Yuen, Meng Zhang, U-Xuan Tan}
\thanks{B. P. L. Lau, S. H. Marakkalage, Y. Zhou, N. Ul Hassan, C. Yuen, U-X. Tan are with the Engineering Product Development, Singapore University of Technology and Design, 8 Somapah Rd, Singapore 487372.}
\thanks{N. Ul Hassan is also with Lahore University of Management Sciences (LUMS), Lahore, Pakistan 54792}
\thanks{Meng Zhang is with Southeast University, Nanjing, China 210096}
}

\markboth{Published in Elsevier Information Fusion}%
{Shell \MakeLowercase{\textit{et al.}}: Bare Demo of IEEEtran.cls for IEEE Journals}

\maketitle

\begin{abstract}
The advancement of various research sectors such as Internet of Things (IoT), Machine Learning, Data Mining, Big Data, and Communication Technology has shed some light in transforming an urban city integrating the aforementioned techniques to a commonly known term - Smart City.
With the emergence of smart city, plethora of data sources have been made available for wide variety of applications.
The common technique for handling multiple data sources is data fusion, where it improves data output quality or extracts knowledge from the raw data.
In order to cater evergrowing highly complicated applications, studies in smart city have to utilize data from various sources and evaluate their performance based on multiple aspects.
To this end, we introduce a multi-perspectives classification of the data fusion to evaluate the smart city applications.
Moreover, we applied the proposed multi-perspectives classification to evaluate selected applications in each domain of the smart city.
We conclude the paper by discussing potential future direction and challenges of data fusion integration.
\end{abstract}


\IEEEpeerreviewmaketitle

\section{Introduction}
\label{sec:Introduction}
According to UN estimates~\cite{un2018world}, $68$\% of the world population would be living in cities by 2050. 
Hence, managing the existing resources and infrastructure to cater sustainable urban living conditions for the growing needs of the urban population has become ever more challenging.
Fortunately, the advancement in Information and Communication Technologies (ICT), Internet of Things (IoT), Big Data, Data Mining, and Data Fusion is gradually paving path for the emergence of smart cities~\cite{boulton201118,hollands2008will,nam2011smart}. 
In this paper, we adopt the following definition of smart city~\cite{toppeta2010smart}:   
\begin{displayquote}
	\textit{``A city combining ICT and Web 2.0 technology with other organizational, design and planning efforts to de-materialize and speed up bureaucratic processes and help to identify new, innovative solutions to city management complexity, in order to improve sustainability and livability''}
\end{displayquote}

The integration of aforementioned technologies into various urban domains enables city managers to equip with the necessary information for better planning and resource management. 
Several cities around the world have already been leveraging these technologies to improve the comfort, security, mobility, health, and well-being of their citizens. 
To better evaluate rapid progress and to recognize the efforts of urban planners, smart city ranking systems have been established. 
For instance, IESE cities in motion index~\cite{berrone2018iese} has suggested $83$ indicators to rank $165$ cities over $80$ countries. 
New York, London, and Paris are the top three smart cities in $2018$. 
Smart city projects in New York~\cite{NewYork_:2018} aim to consistently improve the quality of residents' life, reduce the environmental impacts, increase the street light efficiency, and enhance the water quality. 
Meanwhile, the focus of Smart London Projects~\cite{GreaterLondonAuthority_:2018} is to collect city wide data to provide world class connectivity, security, and smarter streets to its residents.
Digital transformation, sustainability, and urbanization for improving citizen services are at the cores of Paris Smart City Projects~\cite{Paris_:2018}. 
The following up of the top smart city list includes Singapore and Tokyo, which are some other notable smart cities in the world.
In Singapore, Smart Nation Project~\cite{Nation_:2018} has been proposed, which includes e-payment systems, smart nation sensor platform, smart urban mobility, and smart community initiatives, with the aim to enhance the national digital identity of its citizens. 
On the other hand, Tokyo~\cite{Government_:2016} aims to become the greenest city in Asia Pacific by improving the transportation and other sectors of their economy. 
Local governments in several Chinese cities~\cite{Limited_DCPS:2018}, such as, Shenzhen, Shanghai, Hangzhou, and Beijing are also shaping up their cities to facilitate economic and social development to build high income smart cities. 
In addition, there are several research institutes and laboratories focusing on developing smart city applications, which are currently leading the worldwide effort in smart  domains. 
These include MIT Senseable Lab~\cite{MIT_:2018}, Future Cities Laboratory~\cite{Zurich_:2018}, SINTEF Smart Cities~\cite{SINTEF_:2018}, SMART~\cite{SMART_:2018}, etc.

\begin{table}[]
	\caption{Literature Review for Data Fusion on Smart City}
	\fontsize{8pt}{8pt}\selectfont
	\label{tbl:1_reviewList}
	\begin{tabular}{@{}p{0.3\linewidth}|p{0.6\linewidth}@{}}
		\toprule
		\textbf{Surveys} & \textbf{Objectives and Topics Covered} \\ \midrule
		Khaledgi et al.~\cite{Khaleghi_If:2013} & Provides insights on the different types of data fusion techniques by exploring their concept, benefits, and challenges. \\ \midrule
		Castanedo~\cite{Castanedo_TSWJ:2013} & Provides an overall view on the different data fusion techniques and methods. The author also reviewed common algorithms such as data association, state estimation, and decision fusion. \\ \midrule
		Alam. et al.~\cite{Alam_IA:2017} & Provides a comprehensive survey on the mathematical model used in data fusion for specific IoT environments.\\ \midrule
		Wang et al.~\cite{Wang_IJoDSaTI:2016}& Proposes an IoT architecture concept to survey on the different sensor data fusion techniques and also provides an overall view on their evaluation framework. \\ \midrule
		Zheng~\cite{zheng2015methodologies}& Discusses about differences on fusing sources and varying techniques for cross domain data fusion. \\ \midrule
		El et al.~\cite{ElFaouzi_IF:2011}& Provides a survey on the intelligent transportation systems, which use data fusion techniques.\\ \midrule
		Esmaeilian et al.~\cite{Esmaeilian_WM:2018} & Provides a throughout study on waste management for smart city aspects with three categories:  (1) infrastructure for the collection of product lifecycle data, (2) new adapting business model, and (3) waste upstream separation techniques.\\ \midrule
		Da Xu et al.~\cite{DaXu_IToii:2014} & Provides an overall view on the current state of the industries for IoT and discusses key enabling technologies such as communication platforms, sensing technologies, and services. \\ \midrule
		Chen et al.~\cite{Chen_EaB:2018} & Reviews the building occupancy estimation and detection techniques while providing a comparison between different sensor types for cost, detection and estimation accuracy, and privacy issues. \\ \midrule
		Qin and Gu~\cite{Qin_PE:2011} & Introduces the data fusion algorithms in IoT domains and data acquisition characteristics. \\ \bottomrule
	\end{tabular}
\end{table}

Nowadays, communication technology is the backbone for the smart city applications as it provides a channel for applications to transfer data effortlessly.
The ongoing quest for novel, more efficient, low-latency, and cost-effective communication technologies and networks, such as, 5G~\cite{Huang2019Iterative,andrews2014will,boccardi2014five}, wireless sensor networks (WSN)~\cite{Erol-Kantarci_IToSG:2011,sreesha2011cognitive,yue2018comprehensive}, Low Power Wide Area Network (LPWAN)~\cite{georgiou2017low,raza2017low}, and Narrow Band IoT (NB-IoT)~\cite{chen2017narrow,wang2017primer} and their integration in smart city projects is also relentless.

These advancement has made many data sources available due to the potential of sensors collecting data with better coverage and power efficiency of the communication platform.
With the large amounts of data becoming readily available in a smart city, data mining techniques~\cite{Hashem_IJoIM:2016,han2011data} are commonly used in the collected data.
It helps in identifying the essential and important data sources in the smart city applications such as monitoring, control, resource management, anomaly detection, etc.
With the availability of parallel data sources in various smart city domains, data fusion techniques that combine multiple data sources, lie at the heart of smart city platform integration. 
The major objectives of data fusion are to address problematic data while enhancing the data reliability and extracting knowledge from multiple data sources.
\revtwo{The existing survey papers related to smart city applications or data fusion classification are summarized in Table~\ref{tbl:1_reviewList}. 
	Majority of these review papers~\cite{Castanedo_TSWJ:2013,Dasarathy_PotI:1997,DurrantWhyte_Tijorr:1988,Steinberg_:2008} strictly focus on one particular smart city domain or one genre of classification perspective.
	In~\cite{Alam_IA:2017}, Alam et al. have conducted a review on data fusion technique based on mathematical model in IoT environment. 
	Alternately in~\cite{Wang_IJoDSaTI:2016}, Wang et al. have described the frameworks of data fusion within the smart city application.
	Interested readers can follow these references for additional technical details.}
However, there is only a handful of limited work to provide a multi-perspectives approach for data fusion problems in smart cities and this literature gap further motivates our study. 

\begin{figure*}[h]
	\includegraphics[width=1.0\textwidth]{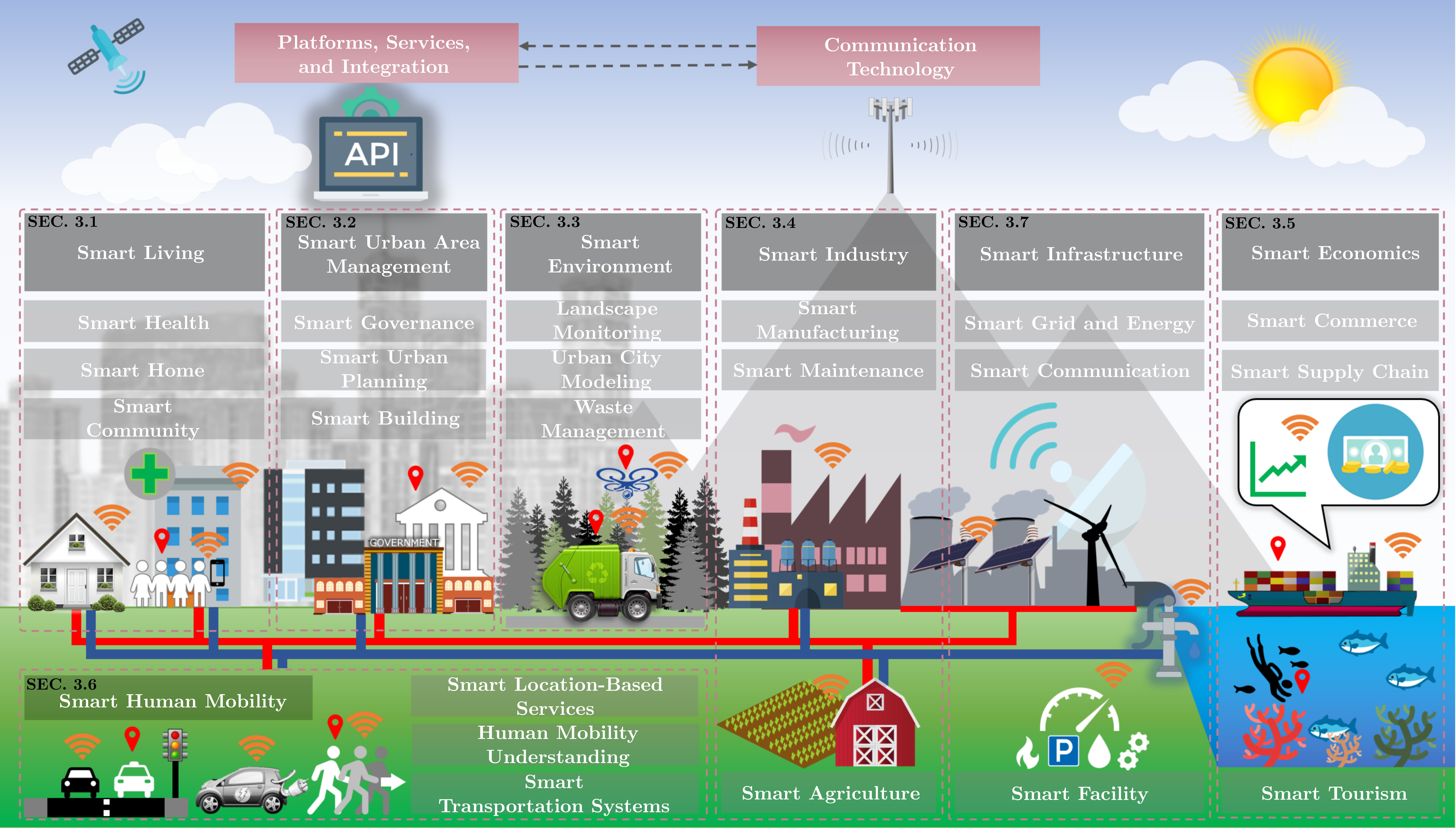} 
	\centering \caption{\revtwo{List of smart city applications domain, where data fusion is commonly applied (each domain is enclosed in the dotted pink box).}}
	\label{fig:Fig01_SmartCityCat}
\end{figure*}

Therefore, a different perspective to look at data fusion in smart city domains is necessitated by the expanding scale and scope of data sources, data collection techniques, and data processing system architectures. 
\revtwo{In order to cater evergrowing highly complicated applications, studies in smart city have to utilize data from various sources and evaluate their performance based on multiple aspects.
	To this end, we propose multiple generic perspectives with the ability to cover the entire depth and breadth of data fusion problems in smart city. 
	These perspectives include data fusion objectives, data fusion techniques, data input and data output types, data sources, data fusion scales, and platform architectures for data processing.
	Utilizing proposed perspectives, we provide an overall view of classification techniques found in the seven domains of smart city applications such as: Smart Living, Smart Urban Area Management, Smart Environment, Smart Industry, Smart Economics, Smart Human Mobility, and Smart Infrastructure.
	A simple illustration of seven application domains discussed in this paper can be found in the Figure~\ref{fig:Fig01_SmartCityCat}.
	In each domain, we only select notable papers to demonstrate the universality and effectiveness of our multi-perspective approach on evaluating the data fusion techniques.
	Please note that we do not provide a comprehensive review of all the smart city applications. 
	Afterwards, we talk about emerging data fusion trends in smart cities, while outlining the best practices for deploying a smart city application. 
	In addition, data fusion challenges in different smart city applications are also identified and discussed. 
	
	To summarize, our novel contributions in this paper are three-fold as shown below:
	\begin{itemize}
		\item We propose a multi-perspectives classification to evaluate common data fusion techniques in smart city applications.
		\item We provide an overview of smart city application domains and discuss the common trend of data fusion techniques in each domain utilizing proposed multi-perspectives classification.
		\item We list down the future challenges and the ideal scenario for deploying data fusion techniques in a smart city application.
	\end{itemize} 
	
	Overall, we believe that with these contributions, the readers would have a quick grasp on the current data fusion trends in smart city research without extensively going through all the details. }

\revtwo{The rest of the paper is organized as follows: in Section~\ref{sec:categoryDefinition}, we define the data fusion classification using multi-perspectives to evaluate a smart city application.
	This lays a foundation for evaluating the smart city applications leveraging data fusion techniques.
	In Section~\ref{sec:currentSmartCityApp}, different application domains of smart city based on data fusion techniques are evaluated using the proposed multi-perspectives classification of data fusion. 
	In addition, a brief overall view of the current research trend of respective domain is presented. 
	Subsequently in Section~\ref{sec:discussion}, we discuss the ideal data fusion scenario along with potential research directions/opportunities based on speculations of smart city applications from previous section. 
	Lastly, we conclude our works in Section~\ref{sec:conclusion}.}


\section{Data Fusion Classification using Multi-perspectives}
\label{sec:categoryDefinition}
In this section, we identify multiple generic perspectives with the ability to cover the entire depth and breadth of data fusion literature in smart city applications. 
We use smart city single perspective data fusion review papers~\cite{Alam_IA:2017,Qin_PE:2011} and non-smart city data fusion classification papers~\cite{Castanedo_TSWJ:2013,Dasarathy_PotI:1997,DurrantWhyte_Tijorr:1988,Steinberg_:2008} as references. 
In non-smart city literature, there are four well-known data fusion classification techniques, which are Dasarathy's Classification~\cite{Dasarathy_PotI:1997}, Whyte's Classification~\cite{DurrantWhyte_Tijorr:1988}, Fusion Architecture's Classification~\cite{Castanedo_TSWJ:2013}, and US Joint Directories of Laboratories (JDL) data fusion classification~\cite{Steinberg_:2008}.
Dasarathy's Classification is based on the data input and output types between data, where Whyte's Classification focuses on the relationship between the data. 
JDL focuses on classifying the fusion process according to five processing levels. 
Meanwhile, the architecture-based classification only captures the system design level and does not consider data relationships and types.
Most of the aforementioned classification of the data fusion techniques are not suitable for evaluating the applications of a smart city. 

Our proposed data fusion classification approach for smart city comprises of six different perspectives (also called categories): i) data fusion objectives ($O$), ii) data fusion techniques ($T$), iii) data input and output types ($D$), iv) data source types ($S$), v) system scales ($L$), and vi) platform architectures ($P$). 
Within each category, we further identify various sub-categories (also called classes). 
Overall, there are $30$ different classes. 
The complete list of the adopted classification indicating all the categories and their classes is shown in Table~\ref{tbl:2_categoryList}. 
Short reference codes ($O$,$T$,$D$,$S$,$L$,$P$) for each class are also included in the table for further use in the paper. 
For example, $O1$ refers to the data fusion objective category and problematic data fusion class. 
Similarly, $S3$ refers to data source types category and participatory class. 

\begin{table}[]
	\caption{Data Fusion Classifications for Smart City Applications using Multi-perspectives }
	\label{tbl:2_categoryList}
	\fontsize{8pt}{8pt}\selectfont
	\begin{tabular}{@{}l|l|l@{}}
		\toprule
		\textbf{Perspective/Category} & \textbf{Code} & \textbf{Classes} \\ \midrule
		\multirow{4}{*}{Data Fusion Objectives} & O1 & Fixing Problematic Data \\
		& O2 & Improving Data Reliability \\
		& O3 & Extracting Higher Level Information \\
		& O4 & Increasing Data Completeness \\ \midrule
		\multirow{9}{*}{Data Fusion Techniques} & T1 & Data Association \\
		& T2 & State Estimation \\
		& T3 & Decision Fusion \\
		& T4 & Classification \\
		& T5 & Prediction / Regression \\
		& T6 & Unsupervised Machine Learning \\
		& T7 & Dimension Reduction \\
		& T8 & Statistical Inference and Analytics \\
		& T9 & Visualization \\ \midrule
		\multirow{5}{*}{Data Input and Output Types} & D1 & Data in Data Out (DAI-DAO) \\
		& D2 & Data In Feature Out (DAI-FEO) \\
		& D3 & Feature in Feature Out (FEI-FEO) \\
		& D4 & Feature in Decision Out (FEI-DEO) \\
		& D5 & Decision in Decision Out (DEI-DEO) \\ \midrule
		\multirow{4}{*}{Data Source Types} & S1 & Physical Data Sources\\
		& S2 & Cyber Data Sources \\
		& S3 & Participatory Data Sources\\
		& S4 & Hybrid Data Sources\\ \midrule
		\multirow{5}{*}{Data Fusion Scales} & L1 & Sensor Level Fusion \\
		& \revtwo{L2} & \revtwo{Building Wide Fusion}\\
		& \revtwo{L3} & \revtwo{Inter-Building Fusion}\\
		& L4 & City Wide Fusion\\
		& \revtwo{L5} & \revtwo{Inter-City Fusion (or Larger)}\\ \midrule
		\multirow{4}{*}{Platform Architectures} & P1 & Edge Computation \\
		& P2 & Fog / Mist Computation \\
		& P3 & Cloud Computation \\
		& P4 & Hybrid Computation \\ \bottomrule
	\end{tabular}
\end{table}

\revtwo{Note that, there could be potentially more than one perspectives (other than data sources, fusion scales, and platform architecture) for smart city application depending on the complexity and fusion objective itself.}
Below, we provide further details of all the perspectives and classes adopted in this paper.

\subsection{Data Fusion Objectives (O)}
The data fusion techniques deployed in a smart city project is influenced by the objective of applications. 
In this paper, we have summarized the four objectives as follows:

\begin{itemize}
	\item \textbf{O1: Fixing Problematic Data} \\
	`Problematic Data' class refers to the case when the data source is having quality issues such as, inconsistency, imperfection, disparateness, etc. 
	Data fusion could be used as an easy approach to overcome such problems. \revtwo{Examples of $O1$ can be found in~\cite{Huang2019Iterative,grime1994data,cheng1997urban,Huang2019Deep}. }
	\item \textbf{O2: Improving Data Reliability} \\
	Data may suffer from reliability issues when it is collected in a less ideal (less controlled) environment with high presence of noise. 
	In such situation, additional data sources are required to add redundancy for increasing data quality to enhance data reliability. 
	\revtwo{Such situations are identified as `Data Reliability' class and~\cite{hong2009evidential,shen2016long,kreibich2014quality,luo2016kernel} exhibits such pattern.}
	\revtwo{In addition, security enhancement through the data fusion also belongs to this category and examples of such objectives can be found in~\cite{li2011communication,petit2015remote,guo2018roboads}.}
	\item \textbf{O3: Extracting Higher Level Information} \\
	Data mining advancement has contributed to many different architectures of data fusion in order to obtain knowledge from multiple data sources.
	For instance, the occupancy of a building can be detected using a combination of few ambient sensors with data fusion, where occupancy information cannot be directly inferred from the raw data sources.
	\revtwo{We classify these approaches as `Higher Level Information Extraction' class and examples can be found in \cite{dawar2018convolutional, jayasinghe2019feature, ghorpade2015integrated}. }
	\item \textbf{O4: Increasing Data Completeness} \\
	In a situation of coverage limitations, an individual data source is insufficient to provide complete details of the desired output.
	\revtwo{Therefore, in `Data Completeness' class, data fusion is performed across multiple data sources to obtain a complete picture of the overall system such as~\cite{luan2010smart,consoli2015urban,ricci2002travel}.} 
\end{itemize}

\subsection{Data Fusion Techniques (T)}
In this category, we present the data fusion techniques in two different information enrichment obtained after data fusion.
The $T1$ until $T3$ are the common data fusion techniques and the further details can be found in \cite{Dasarathy_PotI:1997,Alam_IA:2017}, where it describes the lower level information being fused to generate identical level of information. 
The techniques $T4$-$T8$ are associated with data mining~\cite{han2011data,kotsiantis2007supervised}, where simple input data from multiple sources is fused to generate higher level information enrichment.
Brief description of these classes is given below: 
\begin{itemize}
	\item \textbf{T1: Data Association } \\
	Data association refers to data fusion technique that fuse data based on similarity between at least two or more data sources. 
	\revtwo{Common techniques for data association include Nearest Neighbors~\cite{cover1967nearest}, Probabilistic Data Association~\cite{bar2009probabilistic}, and Multiple Hypothesis Test~\cite{shaffer1995multiple}.} 
	\item \textbf{T2: State Estimation} \\ 
	State estimation indicates the usage of multiple data sources to achieve higher sate estimation accuracy. 
	\revtwo{Common techniques under this category are Maximum Likelihood~\cite{myung2003tutorial}, Kalman Filter~\cite{welch1995introduction}, Particle Filter~\cite{ristic2004beyond}, and Covariance Consistency Model~\cite{uhlmann2003covariance}}. 
	\item \textbf{T3: Decision Fusion} \\
	Decision fusion is a technique that is used to fuse the decisions made by various sub-components of a system to achieve a certain overall objective. 
	For instance, a robot can fuse different decisions from the modules to perform an actuation (direction, events, or actions). 
	\revtwo{General techniques include Bayesian inference~\cite{box2011bayesian}, Dempster-Shafer Inference~\cite{wu2002sensor}, and semantic approaches~\cite{herrera2000fusion}}.  
	\item \textbf{T4: Classification} \\
	Classification technique denotes methodology of grouping objects into different classes based on their unique characteristics.
	\revtwo{In-depth details of generic classification techniques can be found in~\cite{han2011data,kotsiantis2007supervised}}.
	\item \textbf{T5: Prediction} \\
	Prediction techniques are used to forecast output based on single or multiple different data sources.
	Note that, this covers simple methods such as regression and as well as complicated methods such as forecast modeling. \revtwo{Examples of such can be found in~\cite{neter1989applied,makhoul1975linear, lork2017many}}
	\item \textbf{T6: Unsupervised Machine Learning} \\
	Unsupervised machine learning tries to automate the knowledge discovery without relying on the data labels. \revtwo{Examples of such methods involves clustering~\cite{jain1999data}, anomaly detection~\cite{liao2013intrusion} and others~\cite{han2011data}.}
	\revtwo{Note that, semi-supervised machine learning approach~\cite{zhu2005semi} is also categorized under this class.}
	\item \textbf{T7: Dimension Reduction} \\
	Dimension reduction refers to the method of reducing data sources' dimensions for features extraction or visualization purposes. 
	\revtwo{Examples of dimension reduction techniques are Principal Component Analysis (PCA)~\cite{jolliffe2011principal}, and others~\cite{han2011data}.}
	The aim is to preserve the characteristic of the data sources while reducing the complexity of processing high dimensional data.
	\item \textbf{T8: Statistical Inference and Analysis} \\
	Statistical inference and analysis is used for outlining certain information along with some common knowledge / hypothesis from the input data sources. \revtwo{Examples of papers using such approaches can be found in~\cite{Zhang_LaUP:2018, miah2017big}}
	\item \textbf{T9: Visualization} \\
	Visualization is a technique used for the presentation of output to the end users via some platform. The end result often requires human intervention.
	\revtwo{Examples of such techniques can be referred to the following papers~\cite{nichol2005modeling,fan2017heterogeneous,ware2012information}.}
\end{itemize}

\subsection{Data Input and Output Types (D)}
Dasarathy's classification~\cite{Dasarathy_PotI:1997} is based on input and output of fusion technique to determine the relation
between input and output data. There are five classes in data input and output perspective. Brief details are given below:
\begin{itemize}
	\item \textbf{D1: Data In Data Out (DAI-DAO)} \\
	Data In Data Out (DAI-DAO) refers to the situation when multiple raw data sources are fused to increase data reliability and the output after fusion is still a raw data.
	\item \textbf{D2: Data In Feature Out (DAI-FEO)} \\
	Data In Feature Out (DAI-FEO) refers to the situation when multiple raw data sources are fused to extract some unique feature of the observed system. The output feature describes certain aspect of the system and it could be further used for more feature extraction or to make certain decisions.
	\item \textbf{D3: Feature In Feature Out (FEI-FEO)} \\
	Feature In Feature Out (FEI-FEO) refers to the situation when multiple unique features from different sensors are combined to generate new features. 
	This class is commonly known as feature fusion. 
	\item \textbf{D4: Feature In Decision Out (FEI-DEO)} \\
	Feature In Decision Out (FEI-DEO) refers to the situation when certain features of the system are fused to make certain decisions, e.g. actuation of various system components.
	\item \textbf{D5: Decision In Decision Out (DEI-DEO)} \\
	Decision In Decision Out (DEI-DEO) refers to the situation when different decision sources (maintenance status, events, etc.) are combined to obtain a final output decision.
\end{itemize}

\subsection{Data Source Types (S)}
There are four types of generic data sources in smart city applications and \revtwo{we categorize them based on the data sources regardless of the communication medium. Details of each category can be found as follows:}
\begin{itemize}
	\item \textbf{S1: Physical Data Sources} \\
	The physical data sources are collected from sensors that are being deployed to capture information of a particular space, area, or even city wide.
	\revtwo{Examples of the physical sensors include temperature~\cite{Lau_:2016}, air quality~\cite{zheng2013u}, camera~\cite{Spinello_:2011}, ultrasonic~\cite{Lee_:2008}, LiDAR~\cite{chen2004fusion}, and etc. }
	\revtwo{Note that, we categorize smart city application based on the data sources rather than the method they are acquired. For instance, a temperature probe in a sensor nodes of a wireless sensor network (WSN) transmits data through gateway to cloud database is considered as physical data source, $S1$. }
	\item \textbf{S2: Cyber Data Sources} \\
	\revtwo{Cyber data sources denote datasets which are commonly obtained from the Internet domain such as social media information~\cite{Zhang_LaUP:2018,suma2017automatic}, web access data~\cite{breur2011data,wang2012multi}, and opinion based datasets~\cite{balazs2016opinion}. 
		Social media information involves major social media platforms such as Twitter, Facebook, LinkedIn, Weibo, and others. Note that, usually the data is acquired through data mining techniques.}
	Meanwhile, the web access data can be obtained from web applications programming interface (API), such as transportation tickets information and online customer records. 
	Apart from that, open datasets refer to data from third party vendors such as telecom operator or a company with readily available data.
	\item \textbf{S3: Participatory Data Sources} \\
	Participatory data sources include \revtwo{ crowdsensing~\cite{guo2015mobile,marakkalage2019understanding} and crowdsourcing~\cite{estelles2012towards, howe2006rise} data contributed by the personal devices, e.g. mobile phones, wearable devices, tablets, etc. of the users in smart city.}
	Users provide the data voluntarily or through some incentive mechanisms.
	\item \textbf{S4: Hybrid Data Sources} \\
	The hybrid data sources include data obtained from \revtwo{mixed data sources~\cite{Aftab_EaB:2017,You_IToKaDE:2018}}, e.g. by combining the participatory and physical sensor data.
	As pointed in~\cite{zheng2015methodologies}, hybrid data sources can achieve more insights as compared to single data sources. 
\end{itemize}

\subsection{Data Fusion Scales (L)}
The scale of data fusion is also an important classification perspective.
Please note that data fusion scale is based on sensor coverage rather than sensor deployment. There are four different classes, which are described below:
\begin{itemize}
	\item \textbf{L1: Sensor Level Fusion} \\ 
	\revtwo{At the sensor scale, data from various physical sensors is fused to form an output such as~\cite{jayasinghe2019feature,serdio2014fault}.}
	For instance, fusion of data collected by various smartphone sensors is an example of data fusion at sensor level.
	\item \textbf{L2: Building Wide Fusion} \\
	\revtwo{At the building wide scale, data sources collected within a premise or building is fused to form an output. For instance, fusion of building energy and building security data to develop a building management system~\cite{Tushar_ISPM:2018,de2006information,park2014wireless} is an example of data fusion at building level. }
	\item \textbf{\revtwo{L3: Inter-Building Fusion}} \\
	\revtwo{In the inter-building scale, the data sources collected over several buildings are fused to form an output, where the scale of deployment normally includes small area. 
		For example, data sources of several buildings within a university are used to generate a particular output is considered as inter-building scale.
		Other examples of this data fusion scale also can be found in~\cite{zhou2018understanding,katoch2018shading}.}
	\item \textbf{L4: City Wide Fusion} \\
	\revtwo{In the case of city wide fusion, data sources that involve whole city's area as input for the data fusion architecture fall under this class such as~\cite{catania2014approch,toole2015path,mounce2003sensor}}. 
	For instance, the study of citizen behavior involves fusion of data gathered in different areas of the city is considered city wide data.
	\item \textbf{\revtwo{L5: Inter-City Fusion (or larger)}} \\
	\revtwo{At the inter-city fusion (or larger) scale, data from large areas involving one or more cities or terrains (mountains, sea, forests, etc.) is fused to form an output.
		Examples of this scale involve comparing one smart city to another city or data of a city outskirts and its surrounding areas. More examples of inter-city fusion (or larger) can be referred to~\cite{cheng1997urban, izumi2018real,Anjomshoaa_IIoTJ:2018}. }
	
\end{itemize}

\subsection{Platform Architectures (P)}
The architecture of computational platform involved in data fusion is another important classification perspective. 
In this category, we identify four generic classes:
\begin{itemize}
	\item \textbf{P1: Edge Computation Platform} \\
	In edge computation platform, data sources are processed and fused at the edge (i.e. very close to the physical location, where data is actually collected). Edge computation devices include micro-controller, computing devices (Raspberry pi), computers, etc. \revtwo{Such architecture can be found in works such as~\cite{serdio2014fault,park2014wireless,katoch2018shading}.}
	With this architecture, communication overheads and latency can be significantly reduced.
	\item \textbf{P2: Fog Computation Platform} \\
	In fog computation platform, data sources are processed and fused at the middle layer, i.e. between the edge and the cloud.
	In this architecture, data is periodically or continuously sampled at the edge (without processing) and is then forwarded to a gateway (that acts as a fog device). At the gateway, computing resources are provided for data processing.
	\revtwo{Both fog computing and edge computing platforms provide similar benefits of offloading computation as shown in~\cite{catania2014approch,tian2016agri,mehmood2015future}.}
	However, fog computing architecture should be preferred when it is difficult to find stable power sources at the edge.
	\item \textbf{P3: Cloud Computation Platform} \\
	In cloud computation platform, data sources are processed and fused in the cloud.
	This is the most common technique practiced by industry and research institutes for processing big data. \revtwo{Examples of this architecture being used are~\cite{consoli2015urban,breur2011data,ahmed2015integrated}.}
	The advantages of cloud computing architecture includes ready access to the data and both online and offline for further processing or fusing. The disadvantages include increased communication overheads and costs. 
	\item \textbf{P4: Hybrid Computation Platform} \\
	\revtwo{In hybrid computation platform, processing is distributed among two or more layers (edge, fog and cloud) as shown in~\cite{izumi2018real,fleury2010svm,hondori2012monitoring}.}
	In this architecture, depending on the available resources or application objectives, some low level data fusion and processing is done at the edge or fog, while high level information is extracted in the cloud. 
\end{itemize}


\section{Smart City Applications Overview}
\label{sec:currentSmartCityApp}
\revtwo{Smart city applications tend to have extremely diverse requirements, which contribute to a large variety of different techniques and requirements as stated previously in Section~2 for different domains. Thus, it is necessary to evaluate the smart city applications from a more generic perspectives rather than one specific perspective.}
\revtwo{In this section, we select smart city applications with data fusion techniques from different domains listed in Figure~\ref{fig:Fig01_SmartCityCat}, and evaluate them based on multi-perspectives from the Section~\ref{sec:categoryDefinition}.}
Note that, there exist some literatures that are cross-disciplinary, which may involve more than one domain. In order to address the cross-disciplinary smart city applications, we have grouped them into their closest relevant domain. 
In each application domain, we outline sub-domains and present works related to data fusion techniques.
Using the proposed data fusion classification based on multi-perspectives, we discuss the common data sources and fusion techniques, along with the current research trends in each domain.

\begin{table*}[]
	\vspace{-0.5cm}
	\caption{List of Smart City Applications using Data Fusion Technique(s)}
	\vspace{-0.1cm}
	\fontsize{7pt}{7pt}\selectfont
	\begin{tabular}{p{0.2\linewidth}|c|c|c|c|c|c|c|p{0.4\linewidth}}
		\hline
		\rowcolor[HTML]{9B9B9B} 
		\textbf{Domain} & \textbf{Sources} & \textbf{O} & \textbf{S} & \textbf{D} & \textbf{T} & \textbf{L} & \textbf{P} & \textbf{Remarks} \\ \hline
		\rowcolor[HTML]{FFFFFF} 
		& \cite{dawar2018convolutional} & 3 & 1 & 2 & 4 & 1  & 4 & Smart Healthcare \\ \cline{2-9} 
		\rowcolor[HTML]{FFFFFF} 
		& \cite{hossain2017smart} & 3 & 1 & 1 & 4 & \revtwo{4} & 4 & Voice Pathology Detection \\ \cline{2-9} 
		\rowcolor[HTML]{FFFFFF} 
		& \cite{medjahed2011pervasive} & 3 & 3 & 4 & 3 & \revtwo{2} & 4 & Smart Home Healthcare Monitoring \\ \cline{2-9} 
		\rowcolor[HTML]{FFFFFF} 
		& \cite{fleury2010svm} & 3 & 1 & 2 & 4 & 1 & 4 & Daily Activity Classification \\ \cline{2-9} 
		\rowcolor[HTML]{FFFFFF} 
		& \cite{hong2009evidential} & 2 & 1 & 2 & 4 & \revtwo{2} & 4 & Smart Home Activity Recognition \\ \cline{2-9} 
		\rowcolor[HTML]{FFFFFF} 
		& \cite{hondori2012monitoring} & 3 & 1 & 3 & 4 & \revtwo{2} & 4 & Tele-Rehabilitation\\ \cline{2-9} 
		\rowcolor[HTML]{FFFFFF} 
		& \cite{zhang2008information} & 4 & 4 & 4 & 3 & \revtwo{2} & 4 &  Smart Home Control System\\ \cline{2-9} 
		\rowcolor[HTML]{FFFFFF} 
		& \cite{fan2017heterogeneous} & 3 & 4 & 3 & 9 & \revtwo{4} & 4 & Intelligent Video Surveillance \\ \cline{2-9} 
		\rowcolor[HTML]{FFFFFF} 
		& \cite{mehmood2017utilearn} & \revtwo{3} & \revtwo{3,4} & \revtwo{4,5} & \revtwo{4,5} & \revtwo{5} & \revtwo{3} & \textcolor{black}{Distance Learning} \\ \cline{2-9} 
		\rowcolor[HTML]{FFFFFF} 
		\multirow{-9}{*}{\begin{tabular}{c}Smart \\ Living\end{tabular}} & \cite{Chan_:2008} & 3 & 2 & 3 & 1 & \revtwo{4} & 3 & Smart Community \\ \hline
		\rowcolor[HTML]{EFEFEF} 
		& \cite{Aftab_EaB:2017} & 4 & 4 & 4 & 5 & \revtwo{2} & 3 & Building Management \\ \cline{2-9} 
		\rowcolor[HTML]{EFEFEF} 
		& \cite{luo2007autonomous} & 4 & 1 & 2 & 2 & 1 & 1 & Fire Detection System \\ \cline{2-9} 
		\rowcolor[HTML]{EFEFEF} 
		& \cite{Janssen_GIQ:2013} & 3 & 3 & 4 & 3 & \revtwo{5} & 3 & Lean Government \\ \cline{2-9} 
		\rowcolor[HTML]{EFEFEF} 
		& \cite{cheng1997urban} & 1 & 1 & 1 & 1 & \revtwo{5} & 1 & Urban Planning with Satellite Images \\ \cline{2-9} 
		\rowcolor[HTML]{EFEFEF} 
		& \cite{You_IToKaDE:2018,Lau_IIoTJ:2018} & 3 & 4 & 1,2 & 8,9 & \revtwo{4} & 3 & Urban Space Utilization Detection \\ \cline{2-9} 
		\rowcolor[HTML]{EFEFEF} 
		& \cite{consoli2015urban} & 4 & 3 & 1 & 9 & 4 & 3 & Fault Reporting Platform \\ \cline{2-9} 
		\rowcolor[HTML]{EFEFEF} 
		\multirow{-7}{*}{\begin{tabular}{c}Smart \\ Urban Area \\ Management\end{tabular}} & \cite{Zhang_LaUP:2018} & 3 & 4 & 3 & 8,9 & \revtwo{5} & 3 & Landscape Rating Systems \\ \hline
		\rowcolor[HTML]{FFFFFF} 
		& \cite{Anjomshoaa_IIoTJ:2018} & 3 & 1 & 1 & 9 & \revtwo{5} & 3 & City Environment Monitoring \\ \cline{2-9} 
		\rowcolor[HTML]{FFFFFF}
		& \cite{nichol2005modeling} & 3 & 1 & 2 & 2,9 & 1 & 1 & City Building Map Modeling \\ \cline{2-9} 
		\rowcolor[HTML]{FFFFFF} 
		& \cite{Lu_RS:2017} & 3 & 4 & 4 & 4 & \revtwo{5} & 1 & Forest Types Classification \\ \cline{2-9} 
		\rowcolor[HTML]{FFFFFF} 
		& \cite{shen2016long} & 2 & 1 & 1 & 1 & \revtwo{5} & 1 & Long Term Landscape Monitoring \\ \cline{2-9} 
		\rowcolor[HTML]{FFFFFF} 
		& \cite{wolter2011multi} & 4 & 1 & 4 & 4 & \revtwo{5} & 1 & Forest Species Classification \\ \cline{2-9} 
		\rowcolor[HTML]{FFFFFF}
		& \cite{chang2017developing} & 3 & 4 & 2 & 4 & 1 & 1 & Waste Water Treatment \\ \cline{2-9} 
		\rowcolor[HTML]{FFFFFF} 
		\multirow{-7}{*}{\begin{tabular}{c}Smart \\ Environment\end{tabular}} & \cite{catania2014approch} & 4 & 1 & 2,3 & 2,9 & \revtwo{4} & 2 & Urban Solid Waste Management \\ \hline
		\rowcolor[HTML]{EFEFEF} 
		& \cite{serdio2014fault,zhang2018engine} & 4 & 1 & 2,4 & 2,4 & 1 & 1 & Fault Detection \\ \cline{2-9} 
		\rowcolor[HTML]{EFEFEF} 
		& \cite{jayasinghe2018temporal,Ghosh_MSaSP:2007} & 3,4 & 1 & 3,4 & 5 & 1 & 1 & Tools Life Prediction\\ \cline{2-9} 
		\rowcolor[HTML]{EFEFEF} 
		& \cite{de2006information} & 2 & 4 & 4,5 & 3 & \revtwo{2} & 1 &  Decision Support in Manufacturing\\ \cline{2-9} 
		\rowcolor[HTML]{EFEFEF} 
		& \cite{guo2018roboads} & \revtwo{2} & \revtwo{1} & \revtwo{2,4} & \revtwo{2,3} & \revtwo{1} & \revtwo{1} & \textcolor{black}{Autonomous Robots and Security} \\ \cline{2-9} 
		\rowcolor[HTML]{EFEFEF} 
		& \cite{huang2016data} & 3 & 1 & 2,4 & 4,7 & 1 & 1 & Seafood Freshness Classification \\ \cline{2-9} 
		\rowcolor[HTML]{EFEFEF} 
		\multirow{-5}{*}{\begin{tabular}{c}Smart \\ Industry\end{tabular}} & \cite{moshou2005plant,khanum2017towards} & 3,4 & 1 & 2,4 & 2,4,5 & \revtwo{1} & 1 & Agriculture Plant Disease Classification\\ \hline
		\rowcolor[HTML]{FFFFFF} 
		& \cite{breur2011data} & 4 & 2,3 & 1,3 & 1,8 & \revtwo{5} & 3 & Customer Profiling \\ \cline{2-9} 
		\rowcolor[HTML]{FFFFFF} 
		& \cite{sato2015design} & 4 & 4 & 1,4 & 5 & \revtwo{5} & 3 & Consumer Awareness \\ \cline{2-9} 
		\rowcolor[HTML]{FFFFFF} 
		& \cite{tian2016agri} & 4 & 4 & 1 & 9 & \revtwo{5} & 2 & Blockchain and Supply Chain\\ \cline{2-9} 
		\rowcolor[HTML]{FFFFFF} 
		& \cite{Pang2015} & 3,4 & 4 & 2,4 & 1,5,8 & \revtwo{5} & 3 & Supply Chain Management\\ \cline{2-9} 
		\rowcolor[HTML]{FFFFFF} 
		& \cite{miah2017big} & 3 & 2,3 & 2,3 & 5,8 & \revtwo{4} & 3 & Tourist Behavior Analysis \\ \cline{2-9} 
		\rowcolor[HTML]{FFFFFF} 
		& \cite{ricci2002travel} & 4 & 4 & 2,3 & 6 & \revtwo{4} & 3 & Travel Recommendation System\\ \cline{2-9} 
		\rowcolor[HTML]{FFFFFF} 
		\multirow{-7}{*}{\begin{tabular}{c}Smart \\ Economics\end{tabular}} &  \cite{viswanath2014smart} & 3 & 1,2,3 & 2 & 1,4 & \revtwo{4} & 1,3 & Tourist Tracking Application  \\ \hline
		\rowcolor[HTML]{EFEFEF}
		& \cite{lin2010energy,wilhelm2016wearable} & 2,3 & 1 & 1 & 5,1 & \revtwo{4} & 3 & Outdoor Positioning \\ \cline{2-9}
		\rowcolor[HTML]{EFEFEF}
		& \cite{liu2017fusing,liu2017cooperative} & 2,4 & 1 & 1 & 1,2 & \revtwo{2} & 3 & Indoor Positioning \\ \cline{2-9}
		\rowcolor[HTML]{EFEFEF}
		& \cite{liebig2017dynamic,teo2016bim} & 4 & 1 & 4 & 5,1 & \revtwo{4,2} & 3 & Location-based Services \\ \cline{2-9}
		\rowcolor[HTML]{EFEFEF}
		& \cite{toole2015path} & 3 & 3 & 2 & 1 & \revtwo{4} & 3 & Obtaining Origin-Destination Matrices \\ \cline{2-9}
		\rowcolor[HTML]{EFEFEF}
		& \cite{ghorpade2015integrated} & 3 & 3 & 2 & 4 & \revtwo{4} & 3 & Identifying Transportation Modes \\ \cline{2-9}
		\rowcolor[HTML]{EFEFEF}
		& \cite{yoshimura2014analysis} & 3 & 3 & 2 & 1 & \revtwo{2} & 3 & Monitoring Visitors Inside a Building \\ \cline{2-9}
		\rowcolor[HTML]{EFEFEF}
		& \cite{ahmed2015integrated} & 4 & 1 & 4 & 3 & \revtwo{4} & 3 & Traffic Signal Controlling \\ \cline{2-9}
		\rowcolor[HTML]{EFEFEF}
		& \cite{poonawala2016singapore} & 3 & 3 & 2 & 1 & \revtwo{4} & 3 & Analyzing Public Transport Services \\ \cline{2-9}
		\rowcolor[HTML]{EFEFEF}
		\multirow{-9}{*}{\begin{tabular}{c}\revtwo{Smart Human}\\ \revtwo{Mobility}\end{tabular}} &  \cite{li2014sensor} & 4 & 1 & 4 & 4 & 1 & 3 & Autonomous Vehicle Controlling  \\ \hline
		\rowcolor[HTML]{FFFFFF} 
		& \cite{luan2010smart,wang2012multi} & 3,4 & 1 & 2,4 & 5 & \revtwo{4,1} & 1 &  Smart Grid and Power Utilities\\ \cline{2-9} 
		\rowcolor[HTML]{FFFFFF} 
		& \cite{katoch2018shading,huang2018data} & 3 & 1,4 & 1 & 4,5 & \revtwo{3} & 1 & Solar Farm \\ \cline{2-9} 
		\rowcolor[HTML]{FFFFFF} 
		& \cite{izumi2018real} & 3 & 3 & 2 & 2 & \revtwo{5} & 4 & Smart Metering \\ \cline{2-9} 
		\rowcolor[HTML]{FFFFFF} 
		& \cite{grime1994data,Huang2019Iterative} & 1,2 & 1 & 1 & 1,2 & 1 & 1,3 & Communication (5G)\\ \cline{2-9} 
		\rowcolor[HTML]{FFFFFF} 
		& \cite{kreibich2014quality,luo2016kernel} & 2 & 1 & 1 & 1,5 & 1 & 1 & Communication (WSN) \\ \cline{2-9} 
		\rowcolor[HTML]{FFFFFF} 
		& \cite{abeywickrama2017rf} & 4 & 1 & 2,3 & 4 & 1 & 1 &  Drone Detection \\ \cline{2-9}  
		\rowcolor[HTML]{FFFFFF} 
		& \cite{salpietro2015park} & 3 & 4 & 1,2 & 2 & \revtwo{4} & 3 &  Smart Parking System \\ \cline{2-9} \rowcolor[HTML]{FFFFFF} 
		& \cite{park2014wireless} & 4 & 1 & 1 & 2 & \revtwo{2} & 1 & Bridge Monitoring Platform\\ \cline{2-9} 
		\rowcolor[HTML]{FFFFFF} 
		\multirow{-9}{*}{\begin{tabular}{c}Smart \\ Infrastructure\end{tabular}} & \cite{mounce2003sensor} & 3 & 1 & 2,3 & 4,5 & \revtwo{4} & 3 & Water Distribution System \\ \hline
	\end{tabular}
\end{table*}
\subsection{Smart Living}
Smart living concerns with the life of the urban citizens and revolves around the concept of improving live-ability in urban area.
\revtwo{In the literature, the general objectives of utilizing the smart living domain involve data being used to extract higher level information or increasing the data completeness. 
	In addition, smart city applications in this domain often leverage the cloud or hybrid platform architecture.}
In this domains, we have studied three different aspects of smart living, namely, (1) Smart Health, (2) Smart Home, and (3) Smart Community.

\subsubsection{Smart Health}
\revtwo{Healthcare is a crucial component in everyday life concerning medical and public practices using devices as defined by Lee and Co-authors\cite{Lee2011SmartHC,muhammed2018ubehealth}}.
The rapid development of technology (e.g. smartphones and their in-built sensing devices such as heart rate sensors) provides more opportunities to adopt technology in healthcare applications pervasively. 
For telehealth application in smart city, Hossain et al.~\cite{hossain2017smart} have used electroencephalographic (EGG) signals and voice to monitor a specific user's health with the support of cloud technology and doctor's advices. 
In~\cite{medjahed2011pervasive}, work has shown to monitor elderly at home based on fuzzy fusion model using behavioral and acoustical environment data. 
Similarly, Noury~\cite{noury2002smart} also monitors the activities and fall detection of elderly through fuzzy logic by fusing accelerometer, vibration, and orientation sensor.
In~\cite{marakkalage2019understanding}, Marakkalage et al have used crowd-sensing data from a smartphone application (location, noise, light, etc.) and introduced sensor fusion based environment classification (SFEC) to profile elderly people for understanding their daily lifestyle.
In addition, Dawar and Kehtarnavaz in~\cite{dawar2018convolutional} have implemented a Convolution Neural Network (CNN) to combine both depth camera and wearable devices to detect the transition of movements to fall. 
Apart from that, Hondori et al.~\cite{hondori2012monitoring} have proposed using sensor fusion between depth images and inertia to perform tele-rehab in the home.
The main challenge occurs in pervasive smart healthcare data fusion is discussed in~\cite{lee2008issues} as the need of a higher accuracy to improve sensing robustness against uncertainty and unreliable integration. 

\subsubsection{Smart Home}
The concept of Smart Homes plays an important role nowadays in contemporary urban areas. According to Jiang et al.~\cite{jiang2004smart}, the definition of a smart home provides the capability of controlling, monitoring, and accessed appliances \& services through implementation of ICT.
There are currently many big players in developing the smart home appliances such as Amazon, Google, Apple, IBM, Intel, Microsoft, Xiaomi, and others. 
The challenge faced by manufacturers are related with service integration and formulating software ontology platform. 
These are necessary for implementing the services through different vendors and allow for a better integration.
Meanwhile in~\cite{zhang2008information}, physical sensors (soil moisture) and cyber (weather, traffic) have been fused to control home appliances such as alarm clock and water sprinkle. 
The study of user daily activity is yet another important aspect to understand urban citizen well-being.
In~\cite{hong2009evidential}, Hong et al. have combined series of life activities to understand the lifestyle pattern depends on the equally weighted sum operation and Dempster-Shafer theory.
Also, similar study on the user daily activity patterns can be found in~\cite{fleury2010svm}.
Combination of house environmental sensor (infrared, door contact, temperature, hygrometry sensor, microphone) and wearable devices (kinematic sensors) using support vector machine (SVM) can be used to identify the user activity patterns.
In addition, the modeling of human behavior in a smart home~\cite{brdiczka2009learning} in order to generate learning situation models have proven the efficiency of context-aware services.
\revtwo{In addition, smart home security is yet another study field for many researchers~\cite{fernandes2016security,komninos2014survey,dorri2017blockchain} due to increased usage of IoT devices in normal household.}
The research challenges is to develop the applications for the smart houses while retaining the privacy and security of the end user.

\subsubsection{Smart Community}
According to Smart Communities Guidebook~\cite{san1997smart}, a smart community is described as ``a geographical area ranging in size from neighborhood to a multi-county region whose residents, organizations, and governing institutions are using information technology to transform their region in significant ways''. 
There is only a handful of cities focus on this aspect as majority are still in the stage of transforming from facility to community welfare.
First world countries such as USA, Canada, Australia, European Union, and Singapore shown in~\cite{lindskog2004smart} have started up initiatives to create smart communities.
Information fusion for smart community video surveillance system is performed in~\cite{fan2017heterogeneous} to aid neighborhood in terms of security. 
The combination of the different modal surveillance camera provides a vast amount of visual information extraction such as video summarization for highlighting certain events.
\revtwo{A distance learning framework is proposed in \cite{mehmood2017utilearn}, which enables personalized learning to cater what is best for each individual user. It uses data fusion to understand user environment and their activities by means of hybrid data sources.}
Real-time community monitoring also helps to prevent emergency situations and it ensures the safety of community citizens. 
A good example for a smart community application in large-scale is the Social Credit System in China~\cite{liang2018constructing}.
It is a state-owned system to collect data from both public (traffic cameras, transit data etc.) and private (online shopping, fitness trackers etc.) data sources to monitor and analyze user behaviour to generate a single "credit score" for each person, which helps in community well-being.
The techniques fuse these data sources and remains a back box to the general public. 
However, the effect on user privacy with the rise of ``data state'' remains a debate for some~\cite{cheung2018rise}.
A mature citizen should be on alert and always responds to any potential threat, while spreading the awareness to build a safer community in the urban city.

\subsection{Smart Urban Area Management}
Smart urban area management denotes the managing of urban area using ICT. Sub-domains in this regime composed of urban planning, governance, and smart buildings. 
For an application to fit into this definition, the minimum scale would be at the building level (e.g. a building management system).
\revtwo{The main trend of data fusion techniques being applied in this domain mostly consists of objectives of extracting higher level information or increasing the data completeness. 
	The end product of data fusion include visualization of information for respective authorities.}

\subsubsection{Smart Governance}
In smart governance, managing a city is considered as a complex task as the integration of different domains and services is proven to be challenging. 
Transparent services integration is an example of why many governance authorities are having difficulties to sort it out. 
It is hard to strike a balance in developing a transparent governance policy with consideration of sensitive information. 
Therefore, there is only limited study materials available to the best of our knowledge.
Janssen and Estevez~\cite{Janssen_GIQ:2013} have proposed a centralized platform for cutting down government staff by shifting existing organization to rely on integration of platforms. 
\revtwo{The disaster response management is also considered as another vital element for a smart city to carry out any potential counter measurements towards disaster as shown in~\cite{alazawi2014smart}. }
Apart from that, urban reporting system~\cite{consoli2015urban} has collected report from the city wide region on the faulty infrastructure so that immediate actions can be taken to remedy the situation. 
It uses cloud technology and focuses on the display of fused data report, which it also describes the location and types of infrastructure.
Example of research challenges is to remove any potential fake report to prevent misuse of the reporting platform.
\revtwo{Another example of smart governance that involves city safety can be found in~\cite{jin2016smart}, where it can act as an emergency aid application (light pulse on emergency through mesh network) while providing energy efficient lighting to urban area.}
Moreover, there are cities also working on governance platform such as New York~\cite{NewYork_:2018}, Singapore~\cite{Nation_:2018}, Tokyo~\cite{Government_:2016}, Oslo~\cite{Oslo_:2018}, and others. 
The potential research opportunity is to propose consensus protocols within the city for better integration of services.

\subsubsection{Smart Urban Planning}
Urban planning plays an important role in developing the city economy by taking account of well-being of the urban residents.
Traditionally in urban planning, aerial photography and statistical data sources (building size, population number, public amenities, etc.) are combined to understand the current development state of the city. 
The downside of such method is data sources frequently lacks of fine details, which resulting the output result is not representative.
To address such issue, Cheng and Toutin~\cite{cheng1997urban} have combined various satellite and aerial images to generate details for the exiting urban structures.
Alternately, low power sensors are capable to provide a larger coverage with lower deployment cost, which give researchers the opportunity to study different points of interest in the urban area. 
In~\cite{You_IToKaDE:2018,Lau_IIoTJ:2018,Lau_:2016}, a bottom up urban planning method is implemented, where sensors are installed in a designated region to capture space utilization.
From the collected data, urban planners can study public space utilization pattern using an integrated portal.
Here, a hybrid processing method is proposed, where the data processing and fusion occur in different stages of data pipeline.
In addition, a large variety of data sources can be used for urban planning such as physical sensors~\cite{sohn2007data}, photography~\cite{Zhang_LaUP:2018,xu2016multimodal}, or hybrid data sources~\cite{chen2004fusion}.
Despite wide variety of data sources, human interpretation is required when it comes to make decision on a proposed urban design. 
The need of full automated planning system would further benefit the urban planners to combine different data sources in order to achieve a more ideal city planning.

\subsubsection{Smart Building}
Urban building management provides building owner a platform to understand building's energy consumption rate while automating building resources management. 
It has been extensively studied in~\cite{Chen_EaB:2018,raza2015review,baetens2010properties,li2017optimizing} and the current trend is to optimize the building resources such as hot water systems, electrical consumption, and heating ventilation \& air conditioning (HVAC).
In~\cite{Aftab_EaB:2017}, Aftab et al. have combined four different parameters to predict building occupancy to control HVAC using low-cost embedded systems. 
Some other works such as~\cite{Tushar_ISPM:2018,McKenna_EaB:2015,Chen_:2009} also have the same objectives but using different types of data sources. 
The potential solution for better building management system is to rely on fusing weather, human feedback, and electricity price to fine tune the building resources in order to maximize human comfort, while minimizing the energy consumption.
Apart from that, fire alarm system is considered another important features of the smart building management system.
Luo and Su~\cite{luo2007autonomous} have fused three different data sources (flame, smoke, and temperature sensor) to detect any potential fire outbreak and reduce false alarms.
In addition, a notification-based system is implemented to notify the property owner and manager in case of emergency.
In future, potential building safety features may include a group of robots to deal with fire hazards and double duty as building security patrols.

\subsection{Smart Environment}
Smart environment studies the surrounding of a given area of interest, which covers the internal and external surrounding of a city.
\revtwo{From the literature, we observed that majority of the data sources consist of physical and hybrid data sources, while the data scale often represent a large spatial coverage.}
Nowadays, the most common surrounding effects studied in the smart city include urban heat island (UHI), green house effect, and global warming. 
In addition, we have grouped urban waste management under this domain because it also has an environmental impact.

\subsubsection{Landscape Monitoring}
The main challenge of landscape monitoring in smart city is the sensing coverage of the data sources.
To address such issue, two different sensing approaches have been used such as relying on mobile sensing or satellite-based data. 
Mobile sensing~\cite{Anjomshoaa_IIoTJ:2018,cardone2014participact} offers greater sensing capability by leveraging the mobility of moving objects (vehicles or humans). 
The mobile sensing technique provides a large spatial coverage, but it is not suitable for real-time applications unless there are multiple data sources to compensate the lack of spatial resolution concurrently.
The output type of this mobile sensing includes combination of different spatial data in order to complete the data sources before proceed to data processing stage.
Mobile sensing works such as~\cite{zheng2013u,antonic2014urban} utilized different data sources to complete spatial resolution and visualized the ambient changes across the city.
The common characteristic of aforementioned works is feature extraction, which they visualize the processed features from the raw data sources.
Majority of data input and output types in this domain are DAI-DAO and DAI-FEO since physical sensors are the common data sources.
Using the satellite-based data sources, Shen et al.~\cite{shen2016long} have studied the UHI effect in a city using data sources collected over $26$ years.
The UHI index changes are measured through the combination of Landsat and MODIS images data.
Mobile sensing offers a lower deployment cost, where it sacrifice the spatial resolution given there is limited number of sensors. 
Also, it has a lower coverage compared to satellite data sources. 
In contrast, satellite data has a wider coverage of spatial resolution but it frequently needs data enhancement and lacks of finer details.

\subsubsection{Urban City Modeling}
The surrounding natural resources of an urban city such as mountains and forests are considered as important assets of a city.
The most common data sources in modeling the city area are satellite images, which as stated before, it requires data enhancement such as \cite{tu2001new,wu2016improved} before using it.
Therefore, prior work of data fusion ~\cite{Zeng_:2010} was focused on improving the satellite images quality.
Only until recently, the emergence of machine learning algorithms and faster computers have created new ways to extract large variety of satellite image features. 
For instance in~\cite{Lu_RS:2017} and~\cite{wolter2011multi}, forest types classification have been conducted in order to understand the variety of tree species in a specific region of interest.
Both methods involve region-wide data sources and classification techniques, which are used to identify the tree species based on the forest types.
With a lower deployment cost, small satellite (smallsat) and nano satellite (nanosat) could improve spatial coverage to generate a better data sources.
Smart city applications leveraging satellite data will also beneficial from these deployment.

\subsubsection{Waste Management}
With astonishing rate of garbage being generated daily, waste management for an urban city can be rather challenging. 
Thus, it is essential to handle the waste efficiently to improve on sustainability of a city.
An example of such effort could be found in~\cite{Esmaeilian_WM:2018}, where they have proposed three new aspects of a smart waste management system such as: (1) infrastructure to overlook the overall life cycle of the product, (2) new business models revolving the product life cycle for preventing any waste generation, and (3) intelligent sensor networks for waste management facilities. 
In~\cite{catania2014approch}, Catania and Ventura have combined the proximity reading and weight sensor from garbage bin to estimate the garbage capacity of a typical household.
Afterwards, rubbish categories collected from user mobile devices and garbage trucks are combined to keep track of residential participation in recycling scheme.
On the other hand, waste water treatment helps to manage liquid waste of urban city before discharging to river or reuse.
Chang et al.~\cite{chang2017developing} have combined landsat and MODIS dataset in order to trace the water pollution level of a lake. 
On top of that, a web portal has been deployed to visualize and monitor the water pollution region over the time.
Currently, many researchers are working together to develop an efficient waste management system since there is only limited resources available on earth.
The goal is to adopt the 3R (Reduce, Reuse, and Recycle) concept with the help of ICT to improve city resource sustainability.

\subsection{Smart Industry}
With the upcoming Industry $4.0$ standards~\cite{Wang_CN:2016} touted as the gold standard of the future, various industries have been experiencing transformation with automation and data driven approaches. 
\revtwo{The majority of smart industry applications often leverage data collected from physical sensors while data fusion techniques are often performed at sensor or building level.}
Here, smart industry can be divided into three sub-domains, which are Smart Manufacturing, Smart Maintenance, and Smart Agriculture.

\subsubsection{Smart Manufacturing}
Smart manufacturing denotes the factory that depends on ICT to optimize the manufacturing process by increasing the production throughput. 
In~\cite{de2006information}, De Vin et al. have proposed a simulation tool to test out the management decision support by fusing undisclosed data entries and manufacturing process events. 
Similar to the aforementioned approach, decision based fusion can also be seen in~\cite{groger2012data,lee2003manufacturing}, which combines different machinery sensors data and data warehouse entries. 
The data fusion integration also considers supply chain demand in order to further optimize the manufacturing process.
The challenge in this domain is to develop a self-optimizing manufacturing process while delivering the products to meet the demand of supply chain. 
Therefore, smart manufacturing frequently has a high correlation with the supply chain and attempts to deliver the market needs.
\revtwo{In addition, the robotics usage in the smart manufacturing domain is nothing new.
	Guo et al.~\cite{guo2018roboads} have proposed an anomaly detection to combat potential security aspects in the robots using sensor fusion technique such as state estimation.}

\subsubsection{Smart Maintenance}
The reliability and stability of the equipment and machinery is vital to all the industries to ensure smooth operation in production.
Without the guarantee of smooth operation, any downtime can cost damages to reputation and also loses profit. 
Thus, preventive maintenance has been studied in~\cite{jayasinghe2018temporal,Ghosh_MSaSP:2007,Niu_RESS:2010,schmidt2018cloud} and attempts to predict the remaining useful life (RUL) of a machine accurately.
By accurately predicting the RUL, maintenance can be carried out on time to save cost only when needed.
The common data fusion techniques for predicting RUL are neural network (NN) based model such as CNN and Deep NN (DNN).
Please note that, common data source in this sub-domain is physical data source such as machine states, sensors readings, and related parameters.
Nonetheless on the fault detection domain, machine fault detection can be found in~\cite{serdio2014fault,zhang2018engine}, where they describe the problem of fault diagnosis and apply data fusion techniques to overcome.
State estimation and classification have been used to detect the current state of the machinery.
The data sources share some similarity with the preventive maintenance, where lower level of data information is preferred. 
This yields a faster fault detection when compared to a complex data pipeline.
The research challenge here is to develop a generic and a flexible maintenance system for different scale of applications adhering to the goal of accurate fault detection.

\subsubsection{Smart Agriculture}
In order to produce sustainable food resources in smart city, smart farming~\cite{wolfert2017big,walter2017opinion} has become a trend to meet the food supply demand in a smart city.
There are two different sub-domains in smart farming such as land and sea agriculture. 
In the land agriculture aspect, planting crops using controlled environment has shed some light in fulfilling the city needs of fresh supplies.
However, plant disease remains a potential threat to a highly-dense plantation crop framing. 
In~\cite{moshou2005plant}, Moshou et al. have classified the plant disease infection through Self Organizing Map (SOM) by fusing the spectral reflection and fluorescence imaging data. 
This helps to isolate infected crops while it focuses on the production of healthy plants.
Apart from that, electromagnetic induction sensors, vegetarian index, water stress level, and radiance data are combined in~\cite{de2013field} to better determine the partition of the crop field. 
\revtwo{Similar work also can be found in \cite{khanum2017towards}, where Khanum et al. propose an ontology-based fuzzy logic to classify plant disease.}
The research gaps in this domain involve improving live stock management as well as optimizing smart farm.
On the other hand, sea agriculture is responsible for supplying the seafood supplies in a city. 
Obtaining fresh seafood supplies in an urban city sometimes can be rather difficult due to various factors such as delivery, city location, weather, seasonal pricing, etc.
Therefore, a fresh seafood supply in a city is often not guaranteed.
In order to address such issue, Huang et al.~\cite{huang2016data} have provided a solution by integrating two types of cameras for seafood freshness inspection.
Camera and near infrared spectroscopy are fused through PCA and use NN to classify the freshness index.
The research gaps in this domain involve developing large scale fish breeding and also wide varieties of seafood product such as calm, mussels, abalone, etc.
A potential solution such as smart fish breeding with IoT has been proposed in~\cite{Atlas_:2018}, where it suggests using a moving pod to breed fishes while transporting them to destination in a particular destination simultaneously.

\subsection{Smart Economics}
Smart economics can be defined as the generic commercial activities in an urban city ranging from supply chain, logistic, finance center, to tourism. 
All these activities yield potential commercial value to a city, which it depends on the unilateral or bilateral trading relationship.
In this subsection, we discuss smart economics in three major sub-domains, namely,  (1) Smart Commerce, (2) Smart Supply Chain, and (3) Smart Tourism.

\subsubsection{Smart Commerce}
Today, modern e-commerce platforms use multi modal data sources to reach and better understand their customers. 
This helps e-commerce vendors to give better product recommendations for their customers and it helps customers to make their decisions easily.
Fusing customer data such as mobility, credit card purchases, and social media interactions is commonly used in modern recommender systems. 
In~\cite{breur2011data}, Breur introduced the fusion of customer behavior data and market research data to obtain a holistic picture of the customer. 
Investors can leverage financial data to make investment decisions, as Hassan et al.~\cite{hassan2007fusion} have introduced a fusion model of Hidden Markov Model (HMM), NN, and Genetic Algorithm (GA) for stock market prediction. 
Improving the consumer awareness is conducted in~\cite{sato2015design}, by fusing real world (weather, geographical) and cyber world (Twitter, Facebook) data. 
The proposed system has two levels of fusion, which relies on hierarchical-based processing architecture. 
The data combined bottom level input and fed it into upper level for further processing to achieve its objectives. 

\subsubsection{Smart Supply Chain}
In a smart supply chain, it often involves sources and destination tracking in order to understand the flow / processing of the objects.
As discussed in~\cite{christopher2016logistics}, supply chain management and logistic are the fundamental of modern supplies on fulfilling the needs of an urban city.
For instance in food supply chain, three tiers information fusion framework is proposed in \cite{Pang2015} such as: (1)~to accelerate data processing, (2)~shelf life prediction, and (3)~real-time supply chain planning. 
The proposed hierarchical information fusion architecture (HIFA) includes a process that is intelligently transforming the sensor's data sources into usable decision-making information.
Recently, combination of blockchain technology has paved a new way for revolutionizing the existing supply chain. 
In~\cite{tian2016agri}, Tian has shown the integration of blockchain and supply chain in the agri-food supply application.
It aids consumers to trace the origin of food using Radio Frequency Identification (RFID) along with database or WSN. 
The information also includes food origin to help consumers to identify the brand authenticity and avoids consuming counterfeit products.
The research gap in this sub domain concerns with the implementation of smart supply and it needs the involvement from various commercial organizations. 
The consensus and national regulations are also parts of the critical factors of smart supply implementation.

\subsubsection{Smart Tourism}
The advancement of transportation technology has granted accessibility for the humans to move around the globe with ease.
This phenomenon has caused rapid expansion of the tourism commercial values contributed to a city side income. 
Since then, Internet resources such as travel blogs and recommendation systems have influenced public to venture different locations.
For instance, recommendation system~\cite{ricci2002travel} has been developed to recommend the place to travel based on user's information such as socioeconomic (e.g. age, education, and income) and psychological and cognitive (experience, personality, involvement, and
so forth) groups. 
User choices are used as feedback to further fine-tune the recommendation system using Rocchio's method.
Apart from that, Miah et al.~\cite{miah2017big} have combined social media-generated big data (geo-tagged photos of tourist attraction places) to predict tourist behavioral patterns. 
Alternately, Viswanath et al.~\cite{viswanath2014smart} used a smartphone based mobile application to passively track tourist location data and obtain user ratings for tourist attraction places to better understand the preferences of tourists when they visit tourist attractions.
The potential research development for smart travel is to focus on using a smartphone application for improving travel experience by relying on real-time translation and augmented reality (AR) navigation.

\subsection{\revtwo{Smart Human Mobility}}
\revtwo{Human mobility has been an important research area as commuting and traveling play big roles in modern life. 
	With the help of advanced ICT, plentiful data sources related to human mobility have been collected and accessible to researchers, which yields deeper insights into the nature of human mobility as well as better improvement strategies for transportation systems. 
	\revtwo{Smart human mobility}, therefore, means collecting, managing, and analyzing (fusing) various data sources related to different aspects of residents' movement in order to better understand and improve the way people move. 
	Depending on the purpose of different applications, smart human mobility domain can be further divided into three sub-domains:(1)~Smart Location-Based Services, (2)~Human Mobility Understanding, and (3)~Smart Transportation Systems.}

\subsubsection{\revtwo{Smart Location-Based Services}}
\revtwo{This sub-domain aims to get the accurate position of individuals and further to provide services, such as route planning and navigation, to help them travel efficiently and comfortably, in both outdoor and indoor environments.}
For outdoor positioning, Global Positioning System (GPS) has been the most accurate, reliable and dominant technology since it was allowed for civilian use in 1980s~\cite{parkinson1996global, misra2006global}. 
Less-accurate non-GPS positioning approaches, such as wifi-based localization and cell-tower triangulation, are sometimes used instead of (or together with) GPS, because they consume less energy~\cite{lin2010energy, wilhelm2016wearable}.
For indoor positioning, since GPS does not work well indoors, other positioning approaches have been proposed. 
The data collection technologies used for these approaches mainly include Wi-Fi (WLAN), inertial measurement unit (IMU), RFID tags, Bluetooth, global system for mobile communications (GSM), frequency modulation (FM), and ultra-wide band (UWB)~\cite{yassin2016recent, tariq2017non}. 
Meanwhile, multiple data sources are often fused to achieve more accurate localization results~\cite{liu2017fusing, liu2017cooperative}.
Once accurate locations are obtained, either indoors or outdoors, location-based services (e.g. route planning and navigation) can be provided to end users by fusing the location sequences with other information sources such as geographic information system (GIS) data, real-time traffic data, and \revtwo{user preference data~\cite{liebig2017dynamic, teo2016bim,schlingensiepen2016autonomic, delling2017controlling, han2014building,arfat2017parallel}.}
Since the outdoor positioning and location-based services have been well developed and commercialized, the current research trend in this field is mainly focused on improving the performance (accuracy, deployment cost, and energy cost) of indoor systems and services.

\subsubsection{\revtwo{Human Mobility Understanding}}
Positioning systems not only enable the location-based services for individuals but also provide data sources for further monitoring and understanding human mobility in a larger and more comprehensive scale. 
By aggregating and analyzing (fusing) the location data of residents along with GIS data of the environment, various aspects of human mobility can be monitored and the hidden patterns can be obtained.
As summarized in~\cite{zhou2018understanding}, the most common subjects of monitoring and understanding human mobility include distance and duration distributions~\cite{gonzalez2008understanding}, origin-destination matrices~\cite{toole2015path}, individual activity-based mobility patterns~\cite{jiang2017activity}, transportation mode identification~\cite{ghorpade2015integrated}, and densities and flows within a building (or a cluster of buildings)~\cite{yoshimura2014analysis, prentow2015spatio}.
Results obtained from these subjects provide clues for improving transportation system~\cite{demissie2016inferring}, urban planning~\cite{horner2001embedding}, and communication network~\cite{karamshuk2011human}.
Typical studies in this sub-domain usually fuse one data source of people's movement trajectories with the environment information, such as GIS data of the city or floor plan of a building. 
Although this type of approach has produced much deeper insights compared with traditional approach relied on survey data, there is a trend to fuse multiple data sources related to people's movement and obtain a more comprehensive picture of human mobility~\cite{zhang2014exploring, zhang2015comobile}.
\revtwo{Moreover, social media data sources. such as Tweets, also bring in more information regarding the mobility status in cities due to the combination of spatio-temporal data and descriptive text~\cite{suma2017automatic,alomari2017analysis}.}

\subsubsection{Smart Transportation Systems}
Another large part of smart mobility is the improvement of transportation systems, which mainly comes from three aspects: relieving traffic congestion, improving public transportation, and introducing new transport systems.
To relieve traffic congestion, effective light control plays an important role. While existing light control systems are usually based on hand-crafted rules and do not adjust to the rapid dynamics of traffic flows, intelligent light control approaches have been proposed using different data sources, data fusion techniques, and decision making (optimization and control) algorithms~\cite{ahmed2015integrated, wei2018intellilight}.

Challenges in this aspect mainly come from the implementation of such intelligent light control approaches.
Improvement of the public transportation system is mainly conducted through the network and schedule optimization~\cite{yao2014transit}. 
Although these two topics have been thoroughly discussed in the literature, new insights related to the public transit system \revtwo{(e.g. origin-destination matrices and service level obtained from big data)~\cite{poonawala2016singapore, munizaga2012estimation} and more advanced transport modeling tools enabled by big data~\cite{mehmood2017exploring}} have brought new opportunities.

Even if the existing transportation manner has been optimized, there are still problems that cannot be solved, such as last mile issue and driving accidents. 
Therefore, new transport systems, such as bike sharing systems and autonomous vehicle systems, are introduced. 
Advanced ICT and data fusion techniques are the core of the realization of these systems. 
For a bike sharing system, data fusion and analysis helps to understand how the system works and evaluate different operational strategies~\cite{shaheen2011china, schuijbroek2017inventory}. 
As for the autonomous vehicle system, the control of an autonomous vehicle itself is a complex data fusion process, fusing various data sources about the vehicle and the road by advanced machine learning and control algorithms~\cite{li2014sensor,falcone2007predictive}.
\revtwo{Security plays an important role in the autonomous vehicles deployment to ensure reliability of the autonomous driving. Examples of such techniques can be found in~\cite{petit2015remote,ferdowsi2018robust}.}

\subsection{Smart Infrastructure}
In a smart city, infrastructure aims to provide convenience for the public by supplying resources (electricity, gas, and water) or providing services (public facility or communication systems). 
Here, we outline four different sub-domains for discussion, which are (1) Smart Grid, (2) Smart Energy, (3) Smart Facility, and (4) Smart Communication.

\subsubsection{Smart Grid}
The electrical grid provides an intermediate platform for relaying the electricity from the power plant to residential and industrial area. 
The common goal in this sub-domain is to provide reliable and stable electricity supply with the integration of ICT, which is commonly known as smart grid.
Smart grid has been extensively studied in~\cite{luan2010smart,gao2012survey,kordestani2017data,thirugnanam2018energy} and the goal is to address on load and demand balancing of electricity in a particular area, building, or even household. 
Common technique applied in this sub-domain is forecasting, and example of such application can be found in~\cite{luan2010smart}, which it combines the information received from residential meters and predicts the electricity consumption load. 
Wang et al.~\cite{wang2012multi} have proposed a different approach, where the concept of multi agent systems (MAS) is used to predict building energy consumption by denoting each meter as an agent.
The common goal is to use a higher information extraction technique such as prediction, where it allows grid operators to forecast the grid demand to ensure sufficient electricity load.
Test bed currently is the common method for testing out the smart grid use case and has been studied in~\cite{tushar2016smart}.
\revtwo{Another common research topic is security and reliability of the smart grid system. 
	Li et al.~\cite{li2011communication} have proposed a secure state estimation, which it can be used to address single sensor or multi-sensor scenarios. Similar works addressing smart grid security also can be found in~\cite{huang2012state,liu2015abnormal}.}
On the other hand, advanced metering infrastructure (AMI) has been studied along with the smart grid to ensure the electrical metering is tamper-proof while able to accurately measure energy consumption.
For instance, work in~\cite{mclaughlin2013multi} uses the clustering algorithm to identify energy theft accurately while reducing potential false positives. 
Meanwhile, work in~\cite{izumi2018real} has presented a real-time price estimation by fusing local power and global power consumption to understand real-time electric load of the grid. 
In future, prosumers (producer and consumer) will emerge in the smart grid market and sole distributor paradigm will be no longer valid.
This scenario greatly increase the difficulty of the energy demand and load when accounting the energy as a live market

\subsubsection{Smart Energy}
The search for clean energy resources has been an ongoing effort for many researchers in order to cut down the dependency on the fossil fuels. 
Therefore, the clean energy research direction mostly focuses on renewable energy, which propose to go for a green and less carbon footprint energy producing approach.
Nowadays, the most common renewable energy sources emerged in the market are solar farm~\cite{katoch2018shading,huang2018data} and wind power~\cite{sideratos2007advanced,foley2012current}.
Solar energy is generated based on the conversion of the sunlight into electricity, but the energy harvesting technique suffers from limited energy harvesting time.
Thus, solar irradiance prediction is crucial to ensure maximum energy throughput in the solar farm within the limited time.
Huang et al.~\cite{huang2018data} have proposed to use data driven algorithms such as ABB, SVM, BRT, and Lasso, in which the information from neighboring solar plants are combined to accurately predict the solar irradiance. 
Meanwhile in~\cite{jung2014current}, Jung and Broadwater have implemented a statistical model to fuse wind speed, direction, temperature from forecast station and online measurement to determine the total power output of the wind farm.
Most of the aforementioned methods focus on improving efficiency of the existing energy harvesting methodology. 
Future research on the clean energy relies on various data and energy sources in order to construct a high efficiency energy harvesting model.

\subsubsection{Smart Facility}
Smart facility denotes access of physical facility that provides services to the public such as parking facility, water supply, etc.
The most vital facility in a smart city would be water treatment center as clean water is an important necessity for the urban citizens. 
Any potential leakage or downtime of water supply in a city would be proven troublesome.
Mounce et al.~\cite{mounce2003sensor} propose a water leakage detection using classification technique, which combines all the district water meter data.
Similar concept can be applied on other resources such as gas pipe leakage detection or electricity theft in smart grid.
In the public facility, the wear and tear of structures can be a major issue due to the frequent rate of public usage. 
Hence in~\cite{park2014wireless}, Park et al. have combined multi-metric sensors to estimate the bridge displacement. 
Through this, a rough estimation of the structural health can be determined. 
Alternately, Khoa et al.~\cite{khoa2017smart} have proposed a tensor decomposition approach using the facility data sources in order to understand the facility usage details. 
\revtwo{In addition, the emergence of data centers providing various functionalities to the smart city applications such as~\cite{cioara2019exploiting, li2018ultra,kong2015survey} also one of the focuses for the ongoing efforts of smart city.}
There is also a few domains that is highly correlated with smart facility such as Smart Maintenance and Governance~\cite{consoli2015urban}, where integration of a web portal is used to report potential damages. 

\subsubsection{Smart Communication}
Communication in an urban city remains an essential infrastructure for various application platforms to communicate with each other. 
Not all communication platforms and standards are designed equally as each of them serve different purposes.
Therefore, different standards and protocols to meet varying requirements have been established.
Currently, the upcoming 5G technology~\cite{andrews2014will,rappaport2013millimeter} has promised to bring integration of 5G interface with support for older generation spectrum such as LTE and Wi-Fi in order to provide seamless user experience. 
The common data source in 5G standards is raw signal, and that is the reason why data fusion only happens at the edge level.
For example, Huang et al.~\cite{Huang2019Iterative} and Rappaport et al.~\cite{rappaport2013millimeter} have fused raw signals that are divided through multiple antenna during transmission. 
The receivers will receive multiple signal sources and reconstruct the original information being transferred.
Further discussion of the energy efficient trade-off in wireless communication technology can be found in~\cite{mahapatra2016energy, wang2017survey}.
In the IoT domain, wireless sensor network (WSN) is considered a common communication platform because of its wide coverage and low power consumption. 
WSN is built on top of nodes' network, which is smaller than a wireless ad hoc network.
Hence, multiple nodes can be combined for encoding and decoding the packets received.
Kreibich et al.~\cite{kreibich2014quality} and Luo et al.~\cite{luo2016kernel} have proposed approaches to improve communication between WSN focusing on the communication mechanism between nodes.
The main objective is to focus on the reliability of communication channel while maintaining the coverage (from relay to sink nodes) and also low power consumption.
The research significance of communication is undoubtedly a necessity in smart city as it benefits all domains leveraging communication technology.
The main goal is to design efficient and reliable communication protocols to meet different requirements of applications. 
Alternately, low power communication is yet another goal for IoT in order to achieve long sensing operation.

\section{Challenges and Open Research Directions}
\label{sec:discussion}
\begin{figure*}[h]
	\includegraphics[width=1.0\textwidth]{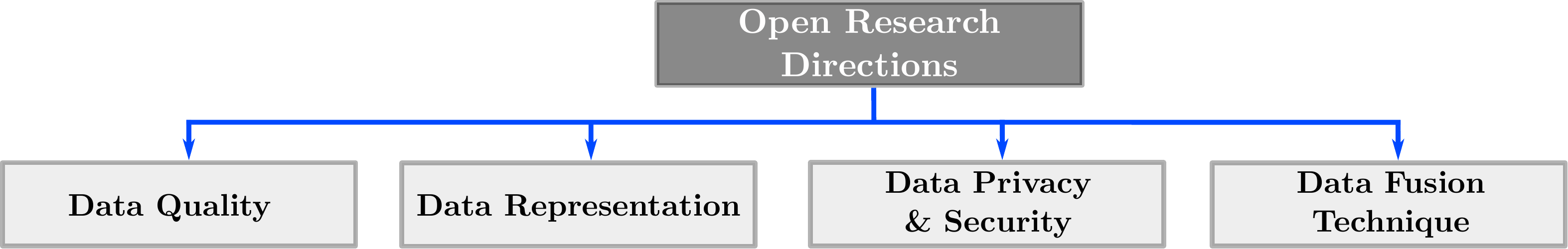} 
	\centering \caption{Open Research Directions for Data Fusion in Smart City Applications.}
	\label{fig:Fig02_open}
\end{figure*}

\revtwo{After outlining the applications of the smart city that use data fusion, we discuss the potential aspects to improve the data fusion in the smart city applications observed from previous section. These aspects include potential categories or perspectives that are not discussed in Section~\ref{sec:categoryDefinition} and~\ref{sec:currentSmartCityApp}.}
As shown in Figure~\ref{fig:Fig02_open}, we identify four major research directions, which are (1)~data quality, (2)~data representation, (3)~data privacy and security, and (4)~data fusion technique.

\subsection{Data Quality}
Quality of the data sources directly determine the quality of output results since processing module follows the ``garbage in and garbage out" theorem in fusing data sources. 
Thus, we discuss two aspects to improve the data sources in the smart city applications, which are sensing coverage and sensing longevity.

\subsubsection{Sensing Coverage}
Sensing coverage is one of the important factors to determine the quality of data sources. 
Insufficient data coverage will generate a result that is not representative, and often it implies more sensors need to be installed to increase the sensing coverage.
This indirectly affects the deployment cost since more physical hardware is required to compensate the sensing coverage.
Apart from that, it also affects the design of communication architecture because more physical sensors are required to transmit data, and thus potentially congests the communication platform.
These factors are common obstacles for a large-scale deployment in smart city applications and getting worse when increasing the deployment scale.
There are two commonly used approaches to address the aforementioned issues, which are crowdsensing and mobile sensing platform.

As shown in~\cite{ma2014opportunities}, crowdsensing is one of the most cost-efficient method as personal mobile devices such as smartphones. 
Smartphones offer wide variety of sensors such as vibration, magnetic field, IMU, GPS, and others. 
The problems with crowdsensing are related to user privacy intrusion and high battery consumption when actively collecting data.
User privacy is a challenge in collecting data as regulations in many countries have been facilitated to prevent applications to collect any sensitive information.
This issue will be further discussed in user privacy and security sub-section.
Another problems with crowdsensing are the unavailability of geolocations information or random distribution of geographical located data.
These scenarios lead to inconsistent data quality.
Potential way to resolve this limitation is to collect data at a fixed time and location only when needed, where incentive is provided for valid participants.
Also, the trade-off problem of the mobile sensing can be further found in~\cite{wang2018distributed}.
Through this method, only qualified data will be included as data sources, while invalid information will be automatically filtered.

Using similar concept as crowdsensing, mobile sensing has offered the same data sensing approach but only follows designated route to collect data. 
The idea is to leverage the mobility of the transportation (normally public transports, cabs, and garbage trucks) to conduct data collection, where the vehicles are traveling across the city. 
Example of mobile sensing platform can be found in~\cite{Anjomshoaa_IIoTJ:2018}, where garbage trucks on duty will collect the ambient data across different parts of the city weekly. 
An identical concept can also be implemented with the public transport systems, since majority of them follow fixed schedules. 
The challenge with the mobile sensing is that spatial resolution of the data may not have a finer detail when compared to crowdsensing due to fixed data collection schedule. 
The main cause is due to the limited accessibility of the vehicles in certain areas (pedestrian path and residential area). 
Potential workaround of this limitation would be combining the mobile sensing and crowdsensing data sources to generate data that covers large area within the urban city. 
Services integration also plays an important role in supplying platforms alternate data sources to perform data enrichment.
By simulating the different IoT services in smart city as shown in~\cite{jha2018holistic}, potential limitation or bottlenecks of smart services can be avoided in order to design a better smart city application.  

\subsubsection{Data Sensing Longevity}
Long term data collection offers different aspects of knowledge discovery as data is able to cover more detail in a larger temporal resolution. 
The advancement of miniaturization has greatly reduced the power consumption of the sensors and IoT devices while maintaining the same sensing performance.
As a result, combining both energy harvesting techniques and low energy devices are able to create a long self-sustaining sensing approach.
This breakthrough allows physical sensors to run independently without the need of external power sources.

In order to preserve the longevity of physical sensors' sensing capability, energy harvesting is one of the common approach in large area networks.
It allows sensors to draw energy from solar energy, vibration, or temperature difference.
The most widely available energy harvesting technique is solar panels and it can be easily obtained. 
Solar panel is affected by the presence of solar irradiance, where the energy harvested varies throughout the different time of the day. 
Contrast to solar farm, the goal here is to conserve as much energy, while maintaining the sensing capability of the physical sensors. 
The most notable influence would be the energy management architecture as well as the battery capacity and the solar panel efficiency.
Apart from that, although temperature difference and sensor vibration are capable of harvesting energy but it is limited to certain use case and not suitable for general usage.

Alternately, potential replacement of the traditional energy harvesting technique is wireless power transfer.
As shown in~\cite{bi2015wireless,sekitani2007large}, this method offers power to be transferred wirelessly without battery and energy harvesting module.
Currently, there are different types of wireless power transfer technologies such as inductive coupling, capacitive coupling, magnetodynamics coupling, microwaves, and light-waves.
Each of them has their limitation such as inductive coupling only has limited range of transferring energy.
That being said, this technology is still relatively new, and it requires further investigation in order to guarantee its minimum working efficiency for smart city applications.

Other than using external power sources, low power sensing for carrying out the sensing tasks.
In order to drive different smart city applications, various standards have been proposed for LPWAN, such as LoRaWAN by LoRa Alliance and NB-IoT Release 13 by 3GPP. 
LoRaWAN focuses on the long range IoT connectivity for industrial applications while the NB-IoT focuses on the indoor coverage, low cost, long battery life, and stable communication in high density communication channel.
The main reason to use low power sensing approach is due to the high compatibility with large scale deployment relying on the low bit rate communication channel usage.
However, standardization of these protocols remains a challenge in LPWAN due to the possibly of using unlicensed spectrum, where organizations may choose not to follow the agreed spectrum. 
In future, low power communication will ensure the long term sensing capability of physical sensors in the smart city applications and therefore will improve data sources quality.

\subsection{Data Representation}
A high speed Internet connection provides easy access to many genres of data sources and creates opportunity to study wide variety of different data sources.
However, large variety of data sources frequently indicate the incompatibility of data formats.
The problem becomes more obvious when there is no standardization on the data format.
To tackle such problem, data ontology is the building block to represent the data sources to connect different sources of data for seamless services integration.
If the format of the data source cannot be interpreted, it will be marked as useless for the platform integrator. 
Therefore, semantic web has been proposed as an extension to WWW web services utilizing Resource Description Framework (RDF) to provide standard data exchange formats.  
It opens the path to create different solutions for the IoT applications and it supports the Open Government Data (OGD) principles~\cite{ubaldi2013open}. 
To date, there are few common ontology languages have been developed such as Delivery Context (DCN)~\cite{cantera2010delivery}, Web Ontology Language (OWL)~\cite{antoniou2004web}, Resource Description Framework Schema (RDFS)~\cite{brickley2000resource}, Semantic Sensor Network (SSN)~\cite{neuhaus2009semantic}, and others.
Majority of the ontology languages only focus on one application domain because they are not suitable for representing the metadata from other domains.
This causes data segmentation in the smart city applications, where further increases the gap between different smart city domains. 
Thus, DBpedia~\cite{auer2007dbpedia} is designed to address the aforementioned issue using public and private stocks of semantic web.
DBpedia has provided solutions for the ontology software as it offers different classes and types that are available on the Wikipedia.
That being said, not all applications adopt the idea of DBpedia and there is a fraction of applications remain conservative using proprietary data representation.
Apart from that, Message Queuing Telemetry Transport (MQTT)~\cite{Organization_:2018} v3.1 protocol has been introduced as one of the protocols to address ontology problems between brokers. 
It offers machine to machine (M2M) communication by providing lightweight publish and subscribe messaging services, where network bandwidth limitation is one of the main constraint. 
It is possible to combine the aforementioned technologies in order to generate a better data integration for data fusion purposes across different domains.
Therefore, the future agenda for the ontology language is to encourage integration of different levels of data sources using different system architecture such as edge, fog, and cloud computing.

\subsection{Privacy and Security}
\subsubsection{\revtwo{Privacy}}
Collecting urban residents' data in a smart city application can be challenging due the nature of sensitive data that can be misused if poorly managed.
As privacy issue has been discussed extensively by the authors in~\cite{martinez2013pursuit,li2016privacy,Chan_:2008}, misuse of private information may lead to catastrophic events such as information theft, or identity fraud.
Currently in Europe, General Data Protection Regulation (GDPR) as discussed in~\cite{tankard2016gdpr,albrecht2016gdpr} has been proposed to better address the data privacy concern of the Internet.
In other countries, there are also similar efforts to enforce data privacy protection such as Canada's Personal Information Protection and Electronic Documents Act (PIPEDA), China's China Data Protection Regulations (CDPR), Singapore's Personal Data Protection Act (PDPA), Japan's Personal Information Protection Commission (PIPC), etc. 
Meanwhile in USA, Health Insurance Portability and Accountability Act of 1996 (HIPAA), the Children's Online Privacy Protection Act of 1998 (COPPA), and the Fair and Accurate Credit Transactions Act of 2003 (FACTA) have been introduced to improve with the information flow efficiency across agencies. 
This is also a part of the efforts to prevent sensitive information being available for unauthorized parties.


Majority of the policies and regulations emphasize on the users' consent for collecting personal data and this can be problematic as not all platforms provide ample security for data storage. 
With insufficient security measurements, the data collected may be compromised, which may lead to tainted reputation and loss of public faith.
For instance, Facebook and Cambridge Analytica scandal~\cite{cadwalladr2018revealed} has shown potential misuse of user data collected. 
With that in mind, potential right of accessing data sources could be revoked if the data source is not handled properly by the right person.
Hence, privacy and security should be the responsibility for both platforms and users. 
A thorough review has been conducted in~\cite{ding2018survey}, which works on the IoT requirements to address privacy issues. 
Potential solution for the aforementioned problem is to use hybrid data fusion technique in a smart city application.
The idea here is to locally fuse the sensitive information (user identity, phone number, bank account number) into generic information, before uploading to the cloud for further processing. 
The benefits of such approach are two-folds, which are the ability to offload computational cost and to preserve sensitive information at the physical sensor only.
In addition, we can leverage machine learning approaches such as~\cite{beaulieu2017privacy,esteban2017real} to generate synthetic datasets with identical data characteristic for study purpose.
This eliminates the chances of private data been leaked out and encourage the openness of datasets to be studied by different researchers and data scientists.
To draw a clear line between generic and sensitive data remains a debate among researchers.
In future, the data fusion can be applied at the lower level to remove any potential sensitive data.

\subsubsection{\revtwo{Security}}
\revtwo{
	According to Kitchin~\cite{kitchin2016getting}, there are two general security concerns in the smart city applications, which are security of technology/infrastructure (data center, services, and system architecture) and data security (data generation, storage, and communication). 
	
	The security of the technology and infrastructure highly relies on the design architecture  of the system being deployed. 
	Depending on the application requirements, it varies from traditional client server architecture to decentralized architecture.
	The main objective is to deploy a hack-proof/exploit-less system architecture. 
	Alternately, there are also ways of improving security of system architecture such as incentive/bounty for reporting flaws, simulating injection attacks, security assessment from third party, etc. 
	Nowadays, the security enhancement focuses towards continuous effort as the technology has been changing rapidly. 
	For instance, security works in~\cite{chakrabarty2016secure, mocanu2019data,talacs2017elastic} have proposed different strategies to enhance the security of the smart city application's architecture by focusing on the common security standards/practices/protocols.
	This shows that as the number of smart city applications increase rapidly, system architectures implemented with the security design in mind become apparent with good practices and standard architecture design. 
	Subsequently, regular security assessment and auditing also pave way for a safer smart city applications deployment.
	
	Meanwhile, data security also contributes to the significant part of smart city applications ecosystem from generation, storage, and communication.
	The common method to combat such issue is leveraging encryption techniques, where it encodes the data so that only the authorized parties have access to it. 
	For instance in~\cite{wang2014performance}, Wang et al. have introduced an attribute based encryption scheme, which it allows fine-grained access control, scalable key management, and flexible data distribution.
	In addition, encryption also can be used in the communication platform between IoT devices in smart city application as shown in~\cite{singh2015secure,elhoseny2018secure} to prevent information hijacking. 
	
	Despite constant effort of cyber security researchers developing new security schemes, the numbers of data breaches and cyber threats increase every year according to David et al.~\cite{DavidMcCandless_:2019}.
	The main culprit of such occurrence is due to negligence of data security practices/implementation. 
	Security often appears to be an afterthought in deployment of a smart city application.  
	Thus, in order to combat such threat, the smart city application should comply with security standards as shown in~\cite{bartoli2011security} to mitigate the chances of becoming a victim.
}

\subsection{Data Fusion Techniques}
Extracting knowledge from a smart city application frequently involves data mining techniques in order to fuse different data sources. 
Lower tier data fusion techniques have been well explored in~\cite{Dasarathy_PotI:1997} and the current research trend focuses more on the machine learning approach. 
The main reason why machine learning approach has gained so much attention is due to its capability of handling high dimensional data.
The problem of high dimensional data is also known as curse of dimensionality as described by Bellman~\cite{bellman2013dynamic}. 
In this context, we discuss two research trends on applying machine learning techniques in data fusion as follows:

\subsubsection{Explainable Deep Neural Network}
Lately, supervised machine learning techniques focus on the DNN, where the in-depth reviews of the recent development can be found in~\cite{zhang2018survey,miikkulainen2019evolving,liu2017survey}.
Major research efforts aim to increase the explainability of the model such as NN, CNN, and DNN rather than using them as black box models.
To this end, explainable AI (XAI)~\cite{Gunning_DARPADnW:2017} is the new motivation for data scientists to explore the interpretable learning paradigm of the modeling in order to provide a semantic meaning behind modeling logic.
This new learning process has driven three big fields in the deep learning domains, which are (1) Deep Explanation, (2) Interpretable Model, and (3) Model Induction.
To develop a deep explanation on the model interpretation, the cognitive layers will act as an intermediate layer between learning and explanation layer in order to cast the learned abstractions, policies, and clusters information into an explainable format. 
Subsequently, the interpretable model such as Bayesian learning~\cite{kendall2017uncertainties} can be built to explain the uncertainties required when developing the deep learning models to learn the choices of a learning process. 
Alternate approach has proposed to use subspace approximation with an adjusted bias technique~\cite{kuo2018interpretable} to build interpretable CNN, which uses feed forward design to better explain the model's choice in allocating certain hyper-parameters.
Meanwhile, model induction refers to the technique used for inferring the model's decision and learning progress.
Through a thorough understanding of the model, parameters can be fine-tuned to increase the learning optimization rate in a long-term application deployment. 
Hence, the search of XAI is an important milestone for the data scientists, which can be used to explain the learning process and the decision machine learning made. 
An example of potential use case would be trying to understand the reason behind (also known as reasoning in some literatures) the predictive maintenance decision machine learning rather than performing maintenance due to the result of predictive algorithm.

\subsubsection{Unsupervised Data Fusion}
In the smart city applications, collecting the ground truth could be proven challenging due to the uncertainties and errors in the collected data sources. 
Hence, obtaining labels or data annotation are another problems with certain data sources.
Despite the rapid development of advanced modeling tools like DNN, it still requires labels and data annotation in order to achieve objectives of extracting higher information.
There are a few approaches that address the lack of labels such as manual annotation, crowd labeling, software annotation, and pattern labeling. 
However, manual annotation only works well with a small dataset while other approaches do not guarantee the correctness of end result. 
This shows a big research gap to seek a better way to label data sources accurately.

Research works such as Zhou et al.~\cite{da2014learning,li2015towards} have attempted to fix unlabeled data by transforming them into useful features to achieve certain objectives.
Traditionally, raw data is required to be preprocessed into something meaningful, but it still suffers from the need of data cleansing and amputation.
The simplest method would be to solely depend on the filtering technique.
However, aggressive filtering may remove large amount of raw data resulting potential loss of knowledge.
Another simple solution is to increase the number of reliable data sources to be fused to create potential annotation.
Increasing data sources often indicates an increment of the overall deployment cost.
Alternative solution to the increased deployment cost is to use transfer learning~\cite{hoo2016deep}, where the knowledge from existing domain can be transferred to other domain to learn from it.

\subsubsection{Emergence of Hybrid Model}
The emergence of the hybrid models has become common due to wide variety of data sources available. 
It allows different levels of data sources  (high, low, or both) to combine in order to create potential insights in a particular domain.
It also helps to solve the data privacy problem along with machine learning technique, which has opened up many opportunities for researchers and data scientist to study on these big data collected.
One example of the hybrid model is shown as follows: an urban planning system has different data sources as input such as human comfort factor index (environmental ambient sensors), positive urban city factor (feedback data on urban area such as greenery, surrounding amenities, recreational parks, and others), and cyber data (social media input) to design a fully automated urban planning system by fulfilling predefined criteria.
The result from the data fusion needs to be explainable as discussed in the previous XAI for understanding choices made by the automation software. 
In this example, different tiers of data sources are fused using data sources types ($D1$, $D2$) and the result is some features.
Eventually, these features will be combined to generate a potential plan for city through computation modeling ($D3$, $D4$). 
By joining different data sources, simulation can be used concurrently to verify the performance of urban planning system before deploying to the city.
In future, implementation of the hybrid model will become a general trend due to wide availability of the data sources and processing platforms.
As mentioned in the discussion, data ontology is another key factor to allow data sources to be connected from different platforms to provide knowledge for the smart city applications.

\section{Conclusion}
\label{sec:conclusion}
This paper presents an overall view of the data fusion techniques found in the smart city applications.
Easy accessibility of the data sources has paved way for data fusion in different smart city applications in various forms.
The increasing trends of data fusion in the smart city applications create the need for a new evaluation method. 
Therefore, we propose a multi-perspectives classification for the smart city applications that involve data fusion techniques.
The data fusion classification based on multi-perspectives introduced in this paper are: (1) Fusion Objectives, (2) Fusion Techniques, (3) Data Input and Output Types, (4) Data Source Types, (5) Data Fusion Scales, and (6) System Architecture. 
Using the proposed multi-perspectives, we evaluated some selected works in the smart city applications and we also discussed the research trend for each domain respectively.
\revtwo{
	Next, we also discuss four open research directions of data fusion in a smart city application such as data quality, data representation, data privacy \& security, and data fusion technique.
	Overall, we are certain that generic nature of the multi-perspectives classification is able to perform well with various smart city applications for different domains that leverage the data fusion techniques.}
\revtwo{In addition, an in-depth analysis can be further extended onto individual domain to study the common requirements and techniques applied, which we do not include in this paper due to limited paper length.
}
A successful smart city application is built on top of the data (also known as data-driven architecture) and data fusion has provided a wide variety of techniques to improve the input data for an application.
Therefore, data fusion has opened the path for various applications to gain insights about the city.
This also holds the key for a smart city to further understand and improve the domains that it is lacking.

\section*{Acknowledgment}
The research work was supported in part by the National Research Foundation (NRF) of Singapore via the Green Buildings Innovation Cluster (GBIC) administered by the Building and Construction Authority (BCA)–Green Building Innovation Cluster (GBIC) Program Office; in part, by the SUTD-MIT International Design Center (IDC; idc@sutd.edu.sg); in part by Natural Science Foundation of China (NSFC) through Project No. 61750110529,61850410535 and Higher Education Commission (HEC) Pakistan through grant number NRPU P\#5913. We thank our colleagues and reviewers, who have provided insight and expertise that greatly assisted with improving the context of this survey paper.

%
%
\bibliographystyle{IEEEtran}
\newcommand{\BIBdecl}{\setlength{\itemsep}{0.25 em}}
\bibliography{bibSpace}

\end{document}